\documentclass[fleqn,usenatbib]{mnras}
\usepackage{newtxtext,newtxmath}
\usepackage{ulem}
\DeclareRobustCommand{\VAN}[3]{#2}
\let\VANthebibliography\thebibliography
\def\thebibliography{\DeclareRobustCommand{\VAN}[3]{##3}\VANthebibliography}
\usepackage{graphicx}	
\usepackage{amsmath}	
\usepackage{soul}
\usepackage{multicol}
\usepackage{cuted}
\usepackage{xcolor}

\newcommand{\CIone}{[C{\small I}] $^3\mathrm{P}_1-^3$$\mathrm{P}_0$}
\newcommand{\CItwo}{[C{\small I}] $^3\mathrm{P}_2-^3$$\mathrm{P}_1$}

\usepackage{animate}
\usepackage{media9}

\usepackage{hyperref}
\usepackage{url}

\title{COALAS II. Extended molecular gas reservoirs are common in a distant, forming galaxy cluster}
\author[Z. Chen et al.]
{
Z. Chen$^{1,2,3,4}$\thanks{E-mail: dg1826004@smail.nju.edu.cn (NJU)},
H. Dannerbauer$^{2,3,}$,
M.~D. Lehnert$^{5}$,
B.~H.~C. Emonts$^{6}$,
Q. Gu, $^{1,4}$,
J.~R. Allison$^{7}$, 
\newauthor
J.~B.~Champagne$^{8}$, 
N. Hatch $^{9}$, 
B.~Inderm\"{u}ehle $^{10}$,
R. P.~Norris$^{10,11}$, 
J.~M. P\'erez-Mart\'inez$^{2,3,12}$,
\newauthor
H.~J.~A. R\"ottgering$^{13}$, 
P. Serra$^{14}$, 
N.~Seymour$^{15}$, 
R.~Shimakawa$^{16}$, 
A.~Thomson$^{17}$,
C.~M. Casey$^{8}$, 
\newauthor
C. De Breuck $^{18}$,
G. Drouart$^{15}$, 
T. Kodama $^{12}$,
Y. Koyama $^{19}$,
C. Lagos Urbina$^{20,21}$, 
P. Macgregor $^{10}$,
\newauthor
G. Miley$^{13}$,
J.~M. Rodr\'iguez-Espinosa $^{22}$, 
M. S\'anchez-Portal$^{23}$,
B. Ziegler$^{24}$
\\
$^{1}$ School of Astronomy and Space Science, Nanjing University, Nanjing 210093, China\\
$^{2}$ Instituto de Astrofísica de Canarias (IAC), E-38205 La Laguna, Tenerife, Spain \\
$^{3}$ Universidad de La Laguna, Dpto. Astrofísica, E-38206 La Laguna, Tenerife, Spain \\
$^{4}$ Key Laboratory of Modern Astronomy and Astrophysics, Nanjing University, Nanjing 210093, China \\
$^{5}$ Université Lyon 1, ENS de Lyon, CNRS UMR5574, Centre de Recherche Astrophysique de Lyon, F-69230 Saint-Genis-Laval, France \\
$^{6}$ National Radio Astronomy Observatory, 520 Edgemont Road, Charlottesville, VA 22903 \\
$^{7}$ First Light Fusion Ltd., Unit 9/10 Oxford Pioneer Park, Mead Road, Yarnton, Kidlington OX5 1QU, UK \\
$^{8}$ The University of Texas at Austin, 2515 Speedway Blvd Stop C1400, Austin, TX 78712, USA \\
$^{9}$ School of Physics and Astronomy, University of Nottingham, University Park, Nottingham NG7 2RD, UK \\
$^{10}$ Australia Telescope National Facility, CSIRO Space $\&$ Astronomy, P.O. Box 76, Epping, NSW 1710, Australia \\
$^{11}$ Western Sydney University, Locked Bag 1797, Penrith, NSW 2751, Australia \\
$^{12}$ Astronomical Institute, Tohoku University, 6-3, Aramaki, Aoba, Sendai, Miyagi, 980-8578, Japan \\
$^{13}$ Leiden Observatory, Leiden University, PO Box 9513, NL-2300 RA Leiden, the Netherlands \\
$^{14}$ INAF- Osservatorio Astronomico di Cagliari, Via della Scienza 5, I-09047 Selargius (CA), Italy\\
$^{15}$ International Centre for Radio Astronomy Research, Curtin University, 1 Turner Avenue, Bentley, WA, 6102, Australia \\
$^{16}$ Waseda Institute for Advanced Study (WIAS), Waseda University, 1-21-1 Nishi Waseda, Shinjuku, Tokyo 169-0051, Japan \\
$^{17}$ Jodrell Bank Centre for Astrophysics, University of Manchester, Oxford Road, Manchester M13 9PL, UK \\
$^{18}$ European Southern Observatory, Karl--Schwarzschild--Stra{\ss}e 2, D-85748 Garching bei M{\"u}nchen, Germany \\
$^{19}$ Subaru Telescope, National Astronomical Observatory of Japan, National Institutes of Natural Sciences, 650 North A’ohoku Place, Hilo, HI 96720, USA \\
$^{20}$ International Centre for Radio Astronomy Research (ICRAR), M468, University of Western Australia, 35 Stirling Hwy, Crawley, WA 6009, Australia \\
$^{21}$ ARC Centre of Excellence for All Sky Astrophysics in 3 Dimensions (ASTRO 3D) \\
$^{22}$ Instituto de Astrofísica de Andalucía, CSIC, Glorieta de la Astronomía, E-18080, Granada, Spain \\
$^{23}$ Instituto de Radioastronom\'ia Milim\'etrica (IRAM), Av. Divina Pastora 7, N\'ucleo Central, E-18012 Granada, Spain \\
$^{24}$ University of Vienna, Department of Astronomy, T\"urkenschanzstrasse 17, A-1180 Vienna \\
}

\date{Accepted XXX. Received YYY; in original form ZZZ}
\pubyear{2023}

\begin{document}
\label{firstpage}
\pagerange{\pageref{firstpage}--\pageref{lastpage}}
\maketitle
\clearpage
\begin{abstract}
This paper presents the results of 475 hours of interferometric observations with the Australia Telescope Compact Array towards the Spiderweb protocluster at \(z=2.16\). We search for large, extended molecular gas reservoirs among 46 previously detected CO(1-0) emitters, employing a customised method we developed. Based on the CO emission images and position-velocity diagrams, as well as the ranking of sources using a binary weighting of six different criteria, we have identified 14 robust and 7 tentative candidates that exhibit large extended molecular gas reservoirs. These extended reservoirs are defined as having sizes greater than 40 kpc or super-galactic scale. This result suggests a high frequency of extended gas reservoirs, comprising at least \(30 \%\) of our CO-selected sample. An environmental study of the candidates is carried out based on N-th nearest neighbour and we find that the large molecular gas reservoirs tend to exist in denser regions. The spatial distribution of our candidates is mainly centred on the core region of the Spiderweb protocluster. The performance and adaptability of our method are discussed. We found 13 (potentially) extended gas reservoirs located in nine galaxy (proto)clusters from the literature. We noticed that large extended molecular gas reservoirs surrounding (normal) star-forming galaxies in protoclusters are rare. This may be attributable to the lack of observations low-J CO transitions and the lack of quantitative analyses of molecular gas morphologies. The large gas reservoirs in the Spiderweb protocluster are a potential source of the intracluster medium seen in low redshift Virgo- or Coma-like galaxy clusters.
\end{abstract}
\begin{keywords}
Galaxy: evolution – galaxies: formation – galaxies: clusters: individual: Spiderweb – galaxies: high-redshift – galaxies: ISM – ISM: molecules
\end{keywords}
\section{Introduction}
Within the paradigm of hierarchical structure formation models, mass is assembled inhomogeneously, along walls, filaments, and nodes all of which constitute the large-scale structure of the Universe or ``cosmic web''. Galaxy protoclusters are the progenitors of the local galaxy clusters, the most massive virialised systems in the Universe with stellar mass \(M \gtrsim 10^{14} M_{\odot}\) (see \citealt{Overzier_2016AARv..24...14O} for review). Galaxy protoclusters residing in the dense regions of the Universe, in which half of the present-day mass is assembled, contribute significantly to the star formation rate density (SFRD) at high redshift~\citep{2017ApJ...844L..23C}. Molecular gas is the direct fuel for star formation in galaxies, thus mapping the molecular gas is essential for studying any potential environmental impact on galaxy evolution in both the field and protoclusters.

Among tens of galaxy protoclusters discovered~\citep{Chiang_2013ApJ...779..127C, Overzier_2016AARv..24...14O}, the Spiderweb protocluster is an excellent laboratory to conduct studies of the molecular content of its member galaxies. Multi-wavelength photometry and spectroscopy data have been accumulated over the last two decades, making the Spiderweb protocluster one of the best studied among all the known protoclusters. Based on the discovery of the clumpy morphology and the surrounding hot, dense magnetised medium from X-ray, optical and radio observations~\citep{Carilli_1997ApJS..109....1C, Carilli_1998ApJ...494L.143C, Pentericci_1998ApJ...504..139P, Pentericci_2000AA...361L..25P}, indications were found that the radio galaxy PKS 1138-262 at z=2.16 is a proto-brightest cluster galaxy (proto-BCG). \citet{Kurk_2000AA...358L...1K} carried out narrow-band imaging of the Ly$\alpha$ line with the Very Large Telescope (VLT) in order to search for a possible overdensity centred around PKS 1138-262 and found 50 Ly$\alpha$ emitters (LAEs) at the same redshift, of which 15 were spectroscopically confirmed in \citet{Pentericci_2000AA...361L..25P} through VLT follow-up observations. A giant Ly$\alpha$ halo ($\sim$ 200 kpc) was discovered, and a few H$\alpha$ emitters (HAEs) are within the halo~\citep{Kurk_2000AA...358L...1K, Kurk_III_2004AA...428..793K, Kurk_IV_2004AA...428..817K}. Based on deep HST imaging~\citep{Miley_2006ApJ...650L..29M}, this radio galaxy is dubbed the ``Spiderweb Galaxy''  and the protocluster associated was later called the ``Spiderweb protocluster''. A panoramic search for HAEs was carried out with MOIRCS/Subaru and revealed  a similar overdensity as LAEs, spanning the scale of $\sim$10 Mpc~\citep{Koyama_2013MNRAS.434..423K, Shimakawa_2014MNRAS.441L...1S, Shimakawa_2018MNRAS.481.5630S}. Near-infrared (NIR) spectroscopy with the VLT/KMOS was carried out for several dozens HAEs in order to study the environmental effect on the ISM through gas-phase metallicity \citep{Perez-Martnez_2022MNRAS.tmp.2575P}. \citet{Rigby_2014MNRAS.437.1882R} found a Herschel-SPIRE overdensity around the Spiderweb Galaxy based on simultaneously conducted 250, 350 and 500 $\mu$m observations. \citet{Dannerbauer_2014AA...570A..55D} complemented this study by revealing an overdensity of intensely star-forming submillimeter galaxies (SMGs) through APEX-LABOCA 870 $\mu$m observations. Through observations with the Australian Telescope Compact Array (ATCA) of the ground transition of carbon monoxide (CO(1-0)) as tracer, large cold molecular reservoirs have been revealed being physically related to the starbursting (proto-)BCG (Spiderweb Galaxy) and the star-forming galaxy HAE229 \citep{Emonts_2016Sci...354.1128E, Dannerbauer_2017AA...608A..48D}. \citet{Jin_2021AA...652A..11J} presented a 21 sq. arcmin panoramic CO(1-0) survey on the Spiderweb protocluster field with the ATCA and reported 46 robust cold molecular gas detections spanning $z=2.09-2.22$, the largest sample of such measurements in a (proto)cluster in the distant universe. The CO emitters are overdense at $z=2.12-2.21$, suggesting a galaxy super-protocluster or a protocluster connected to large-scale filaments with $\sim$120 cMpc size.

There is a growing consensus that the formation of massive galaxies must be a two-phase process, with an early phase driven by gas-accretion, and a late phase dominated by galaxy mergers \citep{Oser_2010ApJ...725.2312O}. The cold molecular gas content and distribution of galaxies, is influenced by various processes, e.g., jet-induced positive feedback, gas accretion and galaxy mergers. All these mechanisms may in principle lead to widespread molecular gas reservoirs. The ATCA has ultra-compact array configurations, which makes it very well suited for detecting low-surface brightness ground-transition CO emission, and thus has played a crucial, leading role in the discovery of extended reservoirs of gas. With the ATCA, a large reservoir of molecular gas ($\sim$ 70 kpc) was found surrounding the Spiderweb radio galaxy, suggesting that the galaxy is growing from the recycled gas in the circum-galactic medium (CGM) rather than through direct accretion from the cosmic web~\citep{Emonts_2016Sci...354.1128E, Emonts_2018MNRAS.477L..60E}. Within the same protocluster, the ATCA detection of an extended ($\sim$40 kpc) rotating molecular gas disk from a normal star-forming galaxy, HAE229, was reported~\citep{Dannerbauer_2017AA...608A..48D}. Besides these two cases within the Spiderweb protocluster, several other large gas reservoir cases have been revealed through various CO transitions and atomic carbon (e.g., \citealt{Ginolfi_2017MNRAS.468.3468G,DAmato2020AA...641L...6D, Umehata_2021ApJ...918...69U, Cicone_2021AA...654L...8C, LiJianrui_2021ApJ...922L..29L}). Very recently, examining the ALMA CO(3-2) observations of seven AGNs (Active Galactic Nuclei) at z$\sim$2-2.5, \citet{Jones_2023MNRAS.518..691J} found evidence for wide-spread ($\sim$ 13 kpc) gas emission.

Extended reservoirs of ionised or molecular gas are considered as one of the characteristics of central galaxies (progenitors of BCGs) in protoclusters~\citep{Overzier_2016AARv..24...14O}. However, the large reservoirs of molecular gas found in the normal star-forming galaxy HAE229~\citep{Dannerbauer_2017AA...608A..48D} indicate that it might not be restricted only to the central galaxy. 

There are multiple processes that likely play a significant role in mass and energy exchange between the interstellar medium (ISM), the circumgalactic medium (CGM), and intergalactic medium \citep[IGM; ][]{Tumlinson_2017ARAA..55..389T}. For example, a galaxy could lose material from the ISM and CGM through AGN or stellar driven outflows, and material may be acquired by a galaxy through gas accretion or recycling the ejected gas within the CGM. The detailed mechanisms involved in regulating galaxy evolution through IGM/CGM/ISM are debated, and the dominant role of each may change depending on the physical scale, mass of the halo, and epoch during which the galaxy is observed. In this paper, the ``extended/large molecular gas reservoirs'' refers to the gas reservoirs with super-galactic scale of tens of kpc ($\gtrsim$ 40 kpc; 40 kpc corresponds to the typical beam-size of the ATCA mosaic data we used in this paper) just like those in the Spiderweb galaxy and HAE229. We note that these gas reservoirs may not necessarily exhibit a smooth distribution, and they can exhibit clumpy structures within this super-galactic scale. Besides the AGNs, which has a violent gas exchange with the surrounding environments (e.g., the Spiderweb galaxy seems to evolve from the recycling of metal-enriched outflow gas; \citealt{Nesvadba_2006ApJ...650..693N}). There are several ways an extended gas reservoirs could originate: (1) retaining of the high angular momentum of the accreting gas; (2) puffing up of the gas in the galaxy by mixing, momentum injection from stars and AGN, and dynamical heating; and (3) galaxy mergers. Filamentary gas accretion/inflow into galaxies is expected in a dense environment like protoclusters, and could be reflected by a gas-phase metallicity gradient. Theoretically, a galaxy accretes gas with a relative small amount of angular momentum, and the gas would inflow into the galaxy centre during the dissipation of angular momentum for a specific timescale (which would further regulate the star-formation \citealt{Lehnert_2015AA...577A.112L}); and the pristine gas accreted along large scale filaments would dilute the central metallicity. \citet{Li_Zihao_2022ApJ...929L...8L} studied gas-phase metallicity gradients of star-forming galaxies in the massive protocluster, BOSS 1244, at redshift z$\sim$2 based on spectroscopic data of the Hubble Space Telescope (HST), and found that the majority of galaxies among their sample have either a negative or flat gradients. They conclude that the cause could be the cold mode accretion ~\citep{Keres_2005MNRAS.363....2K, Dekel_2009Natur.457..451D}, which would flatten or reverse the gradients (the gas-phase metallicity gradient is normally negative, i.e., the metallicity decreases from the inside to the outskirt). Galaxy interactions or mergers may have a similar influence on the gas scales and metallicity gradient of galaxies~\citep{Montuori_2010AA...518A..56M, Rupke_2010ApJ...710L.156R}.

As discussed above, the large molecular gas reservoirs of super-galactic scale are not confined to their ``host'' galaxies but are undergoing complicated processes between the galaxies and their surrounding environment. Revealing the underlying physical mechanism of large gas reservoirs and studying the exchange of such baryonic materials between protocluster galaxies and their environment is essential for understanding galaxy evolution and the origin of gas found in the intracluster medium (ICM) of local clusters and groups. The panoramic COALAS (CO ATCA Legacy Archive of Star-forming galaxies) survey offers a significant opportunity for the first systematic search for large gas reservoirs in a relatively large sample of protocluster galaxies with molecular gas line detections~\citep{Jin_2021AA...652A..11J}. 

It is worth emphasising that the ground transition of carbon monoxide, CO(1-0), is a good tracer of the total molecular gas in star-forming galaxies, including any low-excitation molecular gas that is spread over large scales of tens of kpc (e.g., \citealt{Papadopoulos_2000ApJ...528..626P, Papadopoulos_2001Natur.409...58P, Champagne_2021ApJ...913..110C}). This is because CO(1-0) has a low excitation temperature ($\sim$5 K) and a low critical density ($\rm n_{crit}$) for collisional excitation ($\rm {10}^{3}$ $\rm {cm}^{-3}$) (Table 1 in \citealt{Carilli_2013ARAA..51..105C}). Atomic carbon, such as the transitions \CIone and \CItwo, are used as substitute tracer for CO-dark molecular gas. However, it is less abundant than carbon monoxide (CO), and has higher excitation temperature than CO~\citep{Papadopoulos_2002ApJ...579..270P, Rollig_2006AA...451..917R, Wolfire_2010ApJ...716.1191W, Glover_2012MNRAS.426..377G}. 

In this work, we developed a customised method to search for large, extended gas reservoirs of galaxies within the ATCA COALAS CO(1-0) dataset. Our study is based on the mosaic image presented in~\citet{Jin_2021AA...652A..11J}, which will be referred to as "mosaic" subsequently. Section~\ref{sec:Observation} describes the observations. Section~\ref{sec:DKB03} presents an analysis of the characteristics of a third confirmed large gas reservoir in the Spiderweb protocluster, COALAS-SW.03, utilising higher resolution ATCA observations with an angular resolution of approximately 2-3 arcseconds, equivalent to a physical scale of approximately 17-25 kpc. We compare the morphology and kinematics of COALAS-SW.03 in both the high-resolution and mosaic datasets, where the latter has an angular resolution of approximately 4-5 arcseconds at the location of COALAS-SW.03, equivalent to a physical scale of 34-42 kpc. Our objective is to present a novel methodology for detecting extended molecular gas reservoirs in protoclusters using solely low-resolution data. Section~\ref{sec:sec2_searching} presents the customised method developed to search for extended CO reservoirs (Section~\ref{sec:Method_Head}) and the results of our search (Section~\ref{sec:sec2_results}). A catalogue of 14 robust and seven tentative candidates is provided, and an environmental study is presented. In Section~\ref{sec:discussion}, we then discuss the performance of our method, present a collection of the (possible) large molecular gas reservoirs from the literature, and discuss the underlying environmental physical processes that may create or impact the existence of large molecular gas reservoirs. We assume a flat \(\Lambda\) CDM cosmology with \(\Omega_{\mathrm{m}}=0.3, \Omega_{\Lambda}=0.7\) and \(\mathrm{H}_{0}=70 \mathrm{~km}^{-1} \mathrm{~s}^{-1} \mathrm{Mpc}^{-1}\) in this paper~\citep{2013ApJS..208...19H}. At redshift z=2.16, one arcsec corresponds to 8.4 kpc. 

\section{Observations}
\label{sec:Observation}
\subsection{COALAS CO(1-0) survey}
Several CO observational projects have been conducted with the ATCA on galaxies within the Spiderweb protocluster. Observations include several pointings centred on LABOCA selected sources from \citet{Dannerbauer_2014AA...570A..55D} (ID: 2014OCTS/C3003 and 2016APRS/C3003: PI: H. Dannerbauer), and follow-up high resolution observations centred on the SMG DKB01-03 (ID: 2017APRS/C3003, PI: B. Emonts), the Spiderweb Galaxy, and HAE229 \citep{Emonts_2016Sci...354.1128E, Dannerbauer_2017AA...608A..48D}.

The COALAS (CO ATCA Legacy Archive of Star-forming galaxies) project\footnote{\url{http://research.iac.es/proyecto/COALAS/pages/en/home.php}} is an ATCA large program (C3181; PI: H. Dannerbauer) aiming at studying the environmental impact on the molecular gas based on the CO(1-0) transition. The observations were performed from April 2017 to March 2020, for a total 820 h of observing time. The survey covers both ``cluster'' and ``field'' environments in order to study the environmental effects of galaxy evolution. The two fields are the Spiderweb protocluster ($z = 2.16$) and submillimeter galaxies selected from the Extended Chandra Deep Field-South (ECDFS), respectively. This work focuses on the Spiderweb protocluster and the analysis is based on work presented in \citet{Jin_2021AA...652A..11J}. Combining all these observations, we have 13 pointings (with 475 h of observing time) on the Spiderweb protocluster field. Data reduction was carried out using the software package MIRIAD \citep{1995ASPC...77..433S}. More details of the observations and data reduction can be found in \citet{Jin_2021AA...652A..11J}.

The 13 pointings were placed in order to maximise the number of known sources with spectroscopic redshifts per pointing. Various configurations were used (H75, H168, H214, 750A/C/D, 1.5A; see Table 1 of \citealt{Jin_2021AA...652A..11J} for configurations of each pointing, and the ATCA website \footnote{\href{https://www.narrabri.atnf.csiro.au/operations/array_configurations/configurations.html}{ATCA configurations}} for detailed information). A mosaic datacube was generated by combining 13-pointing observations into a single image (Figure 1 of \citealt{Jin_2021AA...652A..11J}). The mosaic datacube spans an area of 21 square arcminutes and covers a velocity range of $\pm$6500 km/s.  The resulting image has varying levels of noise, ranging from 0.13 to 0.29 mJy root mean square (RMS) in different regions of the image. 

This paper introduces a methodology to effectively identify extended gas reservoirs from the 46 CO detections reported in~\citet{Jin_2021AA...652A..11J}, utilising the 13-pointing mosaic datacube (subsequently referred to as the mosaic/coarse datacube/data). We refer to this datacube as the ``mosaic/coarse datacube/data'', with the term ``coarse'' being interchangeable with ``mosaic''. This distinction is made in comparison to the higher resolution data utilised in the COALAS-SW.03 study. We note that the continuum was separated from the line by fitting a straight line to the line-free channels in the uv domain~\citep{Jin_2021AA...652A..11J}.

\subsection{COALAS-SW.03 observations}
COALAS-SW.03 (dubbed DKB03 in \citealt{Dannerbauer_2014AA...570A..55D}) is one of the CO detections found in \citet{Jin_2021AA...652A..11J}. Following two extended molecular gas reservoirs reported in the literature, the Spiderweb galaxy (COALAS-SW.02 in \citealt{Jin_2021AA...652A..11J}) in \citet{Emonts_2013MNRAS.430.3465E, Emonts_2016Sci...354.1128E, Emonts_2018MNRAS.477L..60E} and HAE229 (COALAS-SW.01 in \citealt{Jin_2021AA...652A..11J}) in \citet{Dannerbauer_2017AA...608A..48D}, COALAS-SW.03/DKB03 is the third large molecular gas reservoir confirmed with high-resolution observations within the Spiderweb protocluster (see Section~\ref{sec:DKB03}).

Our high-resolution CO(1-0) observations of COALAS-SW.03 (ID: 2017APRS/C3003, PI: B. Emonts) were performed with the ATCA during June 2nd to 6th 2017 in the 750m configuration (57.5 h in total). The pointing centre of the observations is on DKB03 (RA: 11:40:57.8; DEC: -26:30:48). Observations were centred around 36.4 GHz, using a channel width of 1 MHz and an effective bandwidth of 2 GHz. Compared to the ATCA 2014OCTS/C3003 and 2016APRS/C3003 observations in H214 configuration, this 2017APRS/C3003 high-resolution observations employed the more extended interferometric array configuration 750m, which provides an angular resolution of $\sim$2-3 arcsec (i.e., $\sim$16.8-25.2 kpc) still guaranteeing sufficient sensitivity in order to detect extended molecular gas emission (i.e., low-surface-brightness emission).

Phase and bandpass calibration were performed by observing the strong calibrator PKS 1124-186, and flux calibration on PKS 1934-638. These data were reduced in MIRIAD~\citep{1995ASPC...77..433S}. Continuum was separated from line-free channels, continuum-subtracted line data generated, binned into 34 km $\rm s^{-1}$ channels and subsequently Hanning-smoothed to an effective velocity resolution of 68 km $\rm s^{-1}$. At the half-power point, the synthesised beam is 4$\farcs$20$\times$1$\farcs$85 (PA = 5.53$^{\circ}$).

\section{COALAS-SW.03: an extended gas reservoir confirmed with high-resolution observations}
\label{sec:DKB03}
In this section, we will characterise COALAS-SW.03 based on the high-resolution ATCA observations in configuration 750m. Then, we compare the morphological and kinematic features of COALAS-SW.03 in both the high-resolution and mosaic/coarse data in order to show the feasibility of searching for large gas reservoirs solely using mosaic data. COALAS-SW.03 was observed at high-resolution using the ATCA 750m configuration, yielding a synthesised beam of 4$\farcs$20$\times$1$\farcs$85. On the other hand, the mosaic data for COALAS-SW.03 is a combination of observations from the H214 and 750m configurations, resulting in a synthesised beam of 4$\farcs$7$\times$4$\farcs$3.

\subsection{Characteristics of COALAS-SW.03}
The molecular gas reservoir of COALAS-SW.03 (dubbed DKB03 in \citealt{Dannerbauer_2014AA...570A..55D}) was first discovered during a sample study of submillimeter galaxies (SMGs) in the Spiderweb protocluster field. Based on the ATCA observations in the H214 configuration (ID: 2014OCTS/C3003 and 2016APRS/C3003: PI: H. Dannerbauer), tentative evidence of the spatial widespread cold molecular gas reservoir ($\sim$ 70 kpc) was found in COALAS-SW.03, and the subsequent high-resolution ATCA imaging in 750m (ID: 2017APR/C3003, PI: B. Emonts) confirmed it. 

\begin{figure*}
    \centering
    \includegraphics[width=1.0\linewidth]{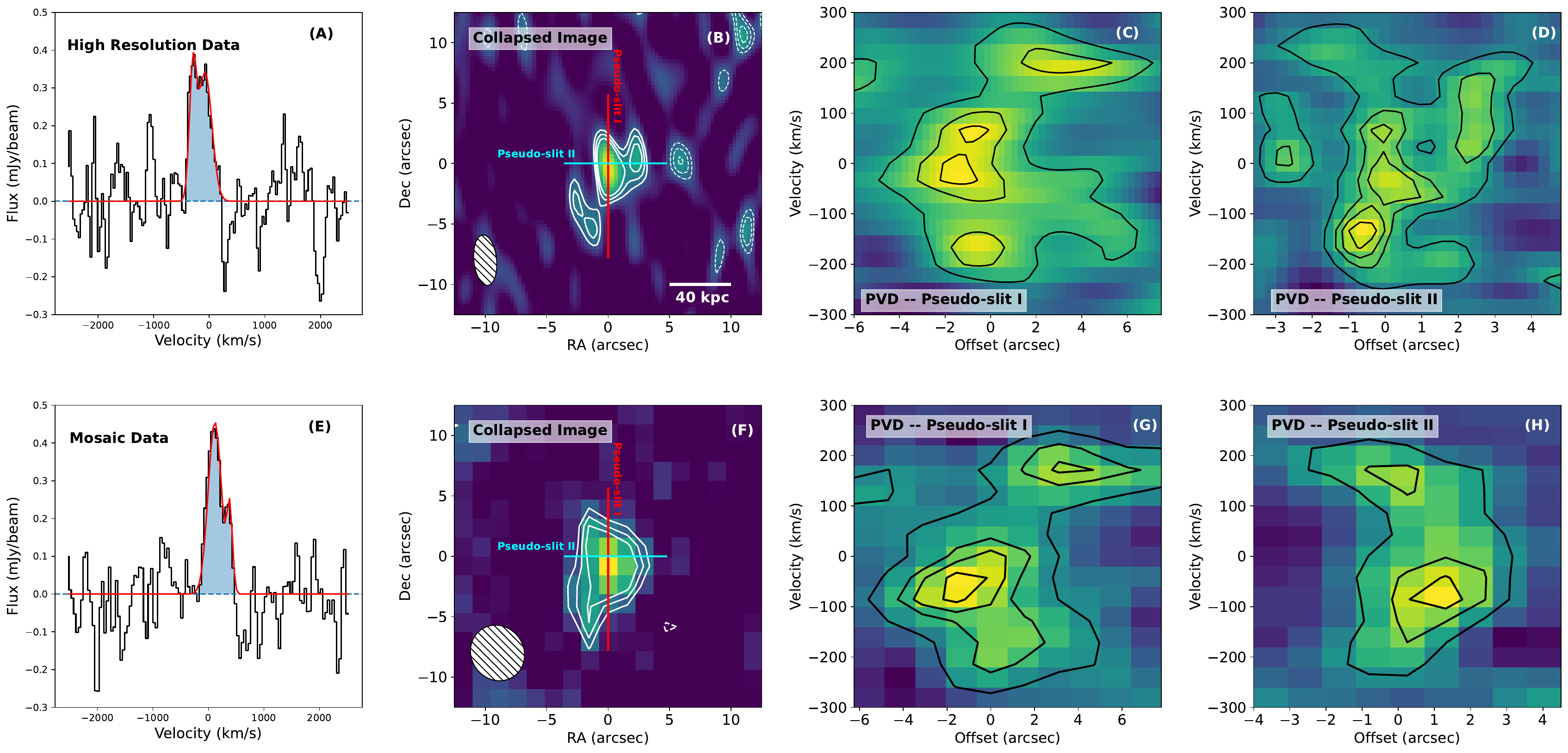}
    \caption{Spectra, collapsed images and PVDs of COALAS-SW.03 (DKB03). The first row shows the characteristics of the high-resolution data, while the second row shows the results from the mosaic datacube previously published in ~\citealt{Jin_2021AA...652A..11J}. Panel (A) and Panel (E) are spectra extracted from the emission peak with aperture sizes of 1$\farcs$5, and the spectral ranges are [-2500, 2500] km/s. A double-Gaussian fitting was initially applied to the spectra to highlight the tentative double-horn features. Panel (B) and Panel (F) are collapsed images (with cut-out sizes of 25$\farcs$0$\times$25$\farcs$0 generated from CO emission line spectral regions shaded in blue in Panel (A) and Panel (E), white contours are [2, 3, 4] $\times\sigma$ ($\sigma$=0.032 Jy $\rm {beam}^{-1}$ km $\rm s^{-1}$), and the synthesised beams are shown at the bottom left of each panel. The white bar at the bottom right of Panel (B) indicates a physical scale of 40 kpc. The RA axis in this and the following figures increases from left to right, which is a departure from the common practice of showing East to the left. However, this orientation choice is not intended to imply that East is to the left. The PVDs shown in panels (C), (D), (G), and (H) are extracted from the red and cyan pseudo-slits shown in panels (B) and (F) for both the mosaic and high-resolution data. With which pseudo-slit the data have been extracted is indicated in the bottom left corner of each PVD. Contours on PVDs are [1.0, 2.0, 2.5, 3.0] times of standard deviation values of each panel.}
    \label{fig:DKB03_Mosaic_AND_HighResolution}
\end{figure*}

The characteristics of the high-resolution observations in the ATCA 750m configurations of COALAS-SW.03 are presented in the first row of Figure~\ref{fig:DKB03_Mosaic_AND_HighResolution}. Panel (A) shows the spectrum  which shows a tentative ``double-horned'' feature. Single-Gaussian fitting results in flux $\rm I_{CO}$ = 0.16$\pm$0.02 Jy km $\rm s^{-1}$, and FWHM=395$\pm$54 km $\rm s^{-1}$. The collapsed image (moment 0) in Panel (B) is generated with the blue shaded spectral range in Panel (A), composed of multiple components. Along the red and cyan pseudo-slits, we extract the position-velocity diagrams (PVDs) of COALAS-SW.03 and display them in Panels (C) and (D). Specifically, we used a slit width corresponding to approximately 5 arcseconds, equivalent to a physical scale of approximately 40 kpc. Notably, careful examination confirmed that slight variations in the slit width had minimal impact on the features presented in Figure~\ref{fig:DKB03_Mosaic_AND_HighResolution}. Both PVDs show multiple velocity components, and a slight velocity gradient is seen along the red pseudo-slit (Panel C). 

\begin{figure*}
    \centering
    \includegraphics[width = 1.0\linewidth]{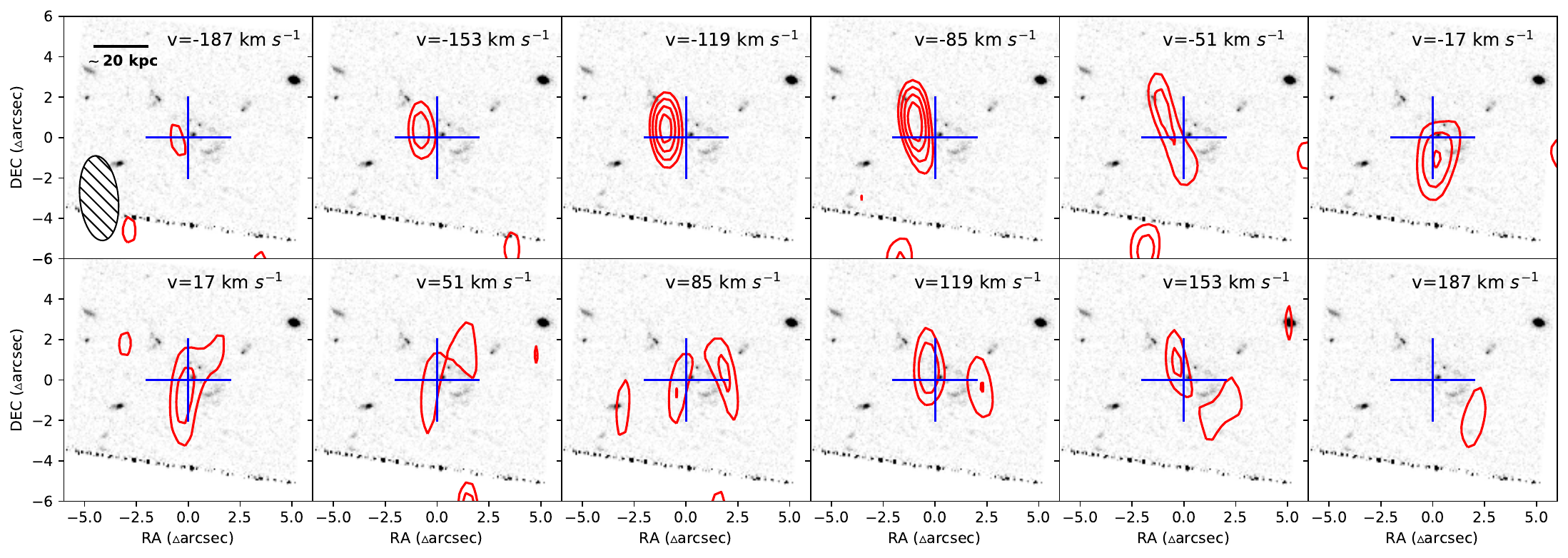}
    \caption{Distribution and kinematics of the CO-emitting gas. High-resolution CO(1-0) maps obtained with the ATCA at different velocities, overlaid on a negative grey-scale HST/ACS F814W image. The velocity centre of each map is indicated in the upper right corner of each panel, and the velocity steps are 34 km $\rm s^{-1}$ between each map. The contours are [2.8, 3.5, 4.2, 4.8, 5.5]$\sigma$. The bottom left ellipse in the first panel shows the synthesised beam (size 4$\farcs$20 $\times$ 1$\farcs$85; position angle: 5.5 deg). The physical scale of 20 kpc (valid for all shown channel maps) is indicated in the upper left corner of the first panel.}
    \label{fig:DKB03_ChannelMap}
\end{figure*}

Channel maps derived from high-resolution observations in Figure~\ref{fig:DKB03_ChannelMap} offer a complementary slice-by-slice view besides the aforementioned moment 0 and PVDs. In Figure~\ref{fig:DKB03_ChannelMap}, we show channel maps for a total velocity range of 374 km $\rm s^{-1}$, with a step in velocity between each map of 34 km $\rm s^{-1}$ (i.e., the velocity range is roughly equal to the FWHM, and the velocity step follows the binning of channels we applied during the data reduction). Multiple components are clearly observed in channel maps, e.g., the panel for a velocity of 85 km $\rm s^{-1}$ shows three separate components.

Either from the projected dimension (collapsed images) or along the line-of-sight (spectrum, PVDs and channel maps), COALAS-SW.03 is an extended clumpy CO emitter composed of multiple components, and the projected size is $\sim$70 kpc.

\subsection{Similarities and differences of COALAS-SW.03 in the high-resolution and Mosaic data}
Comparison between high-resolution data and the mosaic data of COALAS-SW.03 results in some similarities and differences in spectra, collapsed images, and PVDs. Although some detailed features are not resolved in the mosaic data, extended morphologies and multiple components are seen, which is one of the technical foundations of our work on searching for extended large molecular gas reservoirs in the Spiderweb protocluster based on the mosaic datacube (Section~\ref{sec:sec2_searching}). To be more specific, the high-resolution observations of COALAS-SW.03 confirm the extended emission, that was already tentatively seen in the former, coarser data, which means coarse data can be used to find extended sources. Furthermore, the high-resolution data revealed more morphological and kinematics (PVDs) features. However, despite the low resolution, there are multiple-components in PVDs from coarse data and those can be indicators of extended gas reservoirs as well.

Spectra, collapsed images and PVDs derived from the mosaic data are presented in the same format as the high-resolution data in Figure~\ref{fig:DKB03_Mosaic_AND_HighResolution} for comparison.  The detailed comparisons are as follows.

First, the spectrum presented in Panel (E) of Figure~\ref{fig:DKB03_Mosaic_AND_HighResolution} has a similar flux and FWHM as the spectrum extracted from the high-resolution data (with the same aperture). The single-Gaussian fitting results in $\rm I_{CO}$ = 0.16$\pm$0.02 Jy km $\rm s^{-1}$, and FWHM=354$\pm$44 km $\rm s^{-1}$. As we aim to search for large gas reservoirs, we emphasise that the FWHM of CO emission lines also reflect the size of gas detections along the line-of-sight, and the underlying assumption is that a large gas reservoir is supposed to be extended in view of both the projection and along the line-of-sight. 

Second, the collapsed image in Panel (F) is marginally resolved due to the large beam size, which results in different morphology compared to Panel (B). Contours in Panel (B) and (F) are plotted based on standard deviation ($\sigma$) values of each 25$^{\prime\prime}\times25^{\prime\prime}$ cut-out collapsed image shown in Panel (B) and (F) without masking the sources, and the contour levels are [2, 3, 4]$\times\sigma$. The areas encompassed by the outermost contours are similar between the different configurations/resolutions (synthesised beams displayed on the bottom left of Panels B and F).

Finally, the PVDs in Panels (G) and (H) were extracted along the same pseudo-slits as for the high-resolution data. Black contours are based on the standard deviations with levels of [1.0, 2.0, 2.5, 3.0]$\times\sigma$. PVDs of mosaic data show fewer peaks (spatial local maximum) and sub-components than high-resolution data. For PVDs extracted along red pseudo-slits, there are four clear components centred around velocity [-200, -50, 50, 200] km $\rm s^{-1}$ in Panel (C), while only two components are centred around velocity [200, -50] km $\rm s^{-1}$ in Panel (G). The component at velocity 200 km $\rm s^{-1}$ is recovered by mosaic data, the two components of velocity [-50, 50] km $\rm s^{-1}$ are merged into one, and the component at -200 km $\rm s^{-1}$ is not particularly evident at mosaic while being rather an extension of the [-50, 50] km $\rm s^{-1}$ bin. The PVDs extracted along cyan pseudo-slits, due to the much smaller beam size along the west-east direction, Panel (D) reveals more sub-components than Panel (H).

To summarise the similarities and differences of coarse and high-resolution data of COALAS-SW.03 as follows: (1) the high-resolution data recovers well the CO flux and FWHM observed in the mosaic data; (2) the collapsed images, though unresolved in the mosaic data, are of similar isophotal size (comparing the contour encompassed area based on standard value to the synthesised beam area) in both data sets; (3) the PVD components have all a roughly similar morphology, velocity components, and velocity gradient in either data set.  In Section~\ref{sec:sec2_searching}, we present the search for large gas reservoirs mainly based on the spectral FWHM, sizes of collapsed images and features of PVDs.
\subsection{Feasibility of the mosaic Data}
Our comparison between high-resolution and mosaic data demonstrates that the mosaic/coarse data has the potential to uncover large gas reservoirs in terms of both kinematics and morphology. To ensure a robust method of gas reservoir search, we also consider observational conditions for each CO emitter.

The CO mosaic imaging of the Spiderweb protocluster field has inhomogeneities in observational depth and resolution. To accurately assess whether a detection represents a large gas reservoir, we must consider not only its spectral FWHM, collapsed image size, and PVD features (referred to as ``source characteristics''), but also the ``observational conditions'' specific to each detection. This includes the number of pointings covering the source, the configuration and resolution of those pointings, and the positions of the source within each pointing. For a comprehensive explanation of this methodology, see Section~\ref{sec:Method_Head} and refer to Figure~\ref{fig:CriteriaRanking__} for a clear illustration of the criteria related to ``source characteristics'' and ``observational conditions''.

\section{Searching for large molecular gas reservoirs}
\label{sec:sec2_searching}
\subsection{Method}
\label{sec:Method_Head}
Based on the generated mosaic image (Figure 1 in \citealt{Jin_2021AA...652A..11J}), we checked for all these 46 detections their collapsed images with specified locations and velocity channel ranges. Our aim is to develop a generally applicable method for identifying large gas reservoir candidates. The presented method, customised for the COALAS CO(1-0) survey conducted on the Spiderweb protocluster, exhibits universal applicability to other interferometric-based investigations targeting extended gas reservoirs.

\begin{figure}
    \centering
    \includegraphics[width = 1.0\linewidth]{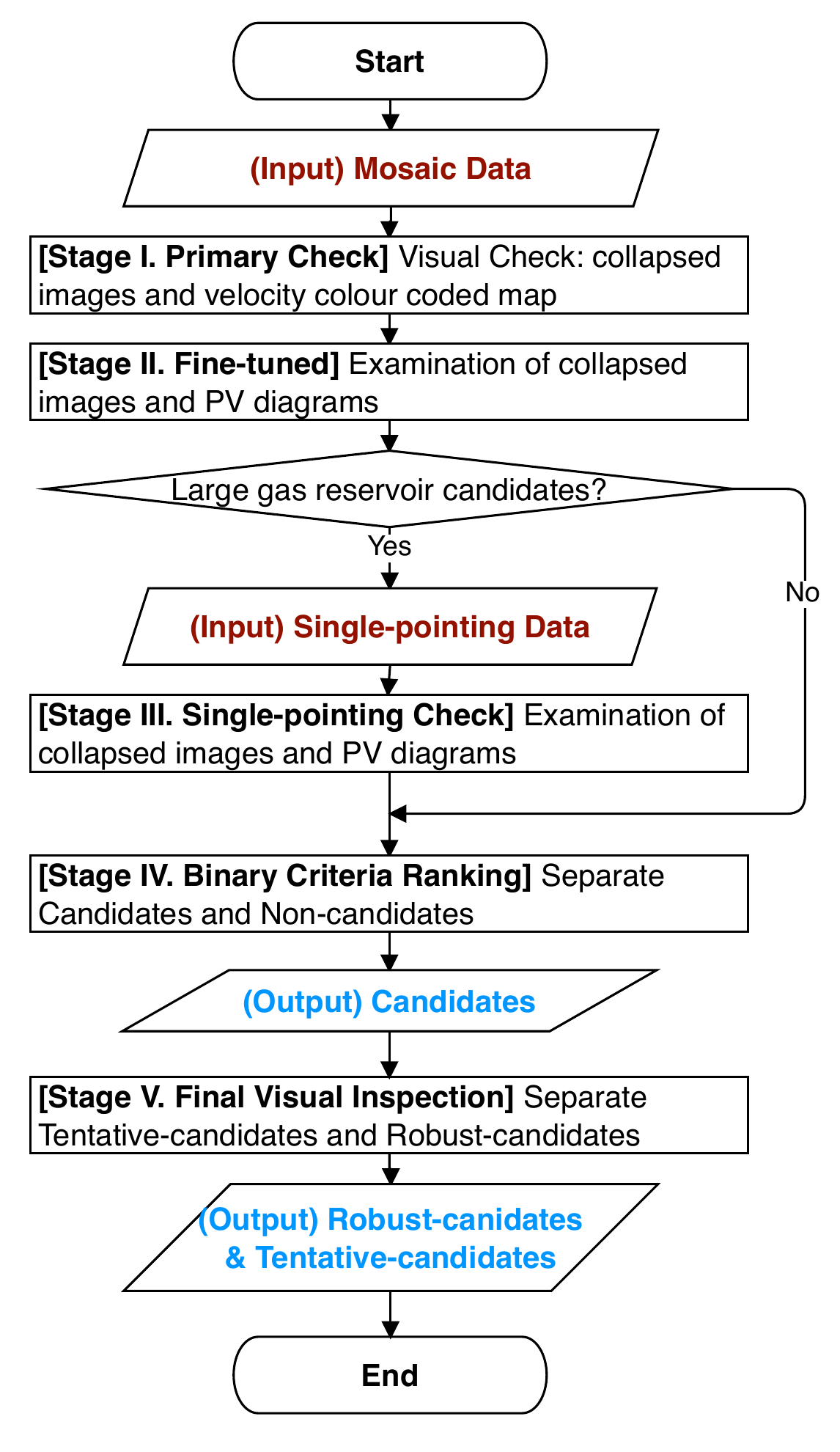}
     \caption{Flowchart of the large molecular gas reservoir search procedure. The procedure is composed by five stages, based mainly on the mosaic data and complemented with the verification of single pointing data cubes for possible candidates selected through the first two stages. }
    \label{fig:FlowChart_1}
\end{figure}

The search process has five stages: (I) primary check; (II) examination of the collapsed images and PVDs obtained from the mosaic imaging data; (III) check for consistencies between the single-pointing and mosaic datasets for all candidates found in the mosaic data set; (IV) priority ranking based on assigning binary numbers to each characteristic; and (V) manual classification of the selected candidates into two classes, i.e., robust and tentative. A flowchart is shown in Figure~\ref{fig:FlowChart_1}. We will describe the procedure of the flowchart in detail in Sec.~\ref{sec:method_I}. In stage IV ``Binary Criteria'' ranking, we designed a set of quantified parameters, reflecting the characteristics of the source and representing the observational conditions. The details of the parameter set-up of stage IV are described in Sec.~\ref{sec:method_II}. The parameter set-up is calibrated and explained in Sec.~\ref{sec:method_IV}.

\subsubsection{Search procedure}
\label{sec:method_I}
The procedure of searching for large molecular reservoirs is shown in the flowchart in Figure~\ref{fig:FlowChart_1}. The detailed explanation is as follows.

\textbf{Stage I. Primary Check:} visual check of collapsed images and velocity colour coded maps. We generated cut-outs (sub-datacube) of 46 sources with various sets of projected spatial  (such as 25$\farcs$, 35$\farcs$, 45$\farcs$) and channel sizes based on CO(1-0) FWHM given in \citet{Jin_2021AA...652A..11J}. 

\textbf{Stage II. Fine-tuned:} the ``fine-tuned'' generation and inspection of collapsed images and PVDs. Based on the first stage, we generated the collapsed images and PVDs for sources more precisely. The spatial and channel sizes were tuned iteratively for each source, ensuring the cut-outs fit the full 3-dimension sizes of sources and minimise the visual influence from the background noise. We constructed a list of potential candidates by comparing the size of collapsed images with observational synthesised beam sizes and visual checking of gradients and multiple components on PVDs. 

\textbf{Stage III. Single-pointing Check:} check the single-pointing data of the candidates we have obtained. Besides the collapsed images and PVDs, we also extracted the spectra from various apertures (single central pixel, 3$\farcs$0 circular aperture, and full spatial cut-out size). 

\textbf{Stage IV. Binary Criteria ranking:} we employ a binary ranking system which assigns a score to each candidate based on multiple criteria. These criteria include: (1) size of collapsed image; (2) PVD features; (3) signal-to-noise ratio of the CO emission detection; (4) number of pointings with small beam sizes ($<$7$\farcs$0); (5) fraction of pointings with small beam sizes; and (6) fraction of pointings in which the source is located near the edge of the primary beam. The first two criteria reflect the characteristics of the source, and the last four account for the observational conditions and the inhomogeneous nature of the data set (i.e., configurations/resolutions and background noise levels). Details can be found in Section~\ref{sec:method_II} and Appendix~\ref{sec:A0_CriteriaRankingDesign}. With this scheme, we are able to rank the 46 CO detections, and separate sources which show evidence for large gas reservoirs from those without (clear) evidence  (Section~\ref{sec:Result_Candidates}).

\textbf{Stage V. Final visual inspection:} We perform a final manual classification of large gas reservoir candidates by visually inspecting the kinematic and morphological features of their collapsed images and PVDs in order to differentiate between robust and tentative candidates.

\subsubsection{Ranking Criteria}
\label{sec:method_II}
We propose a ``Binary Criteria Ranking'' to weight the possibility that a source has an extended/large molecular reservoir. In such a ranking, we selected a set of criteria characterising each source, and place these criteria in order of relevance, prioritising source sizes. The determination of whether a source is extended or not is based on multiple factors, including but not limited to the quality and resolution of the observational data. Consequently, our analysis takes these factors into account, and the established criteria encompasses two distinct aspects: (1) intrinsic features of the source such as the source size and morphology of collapsed images, kinematics as gauged from the PVDs, the spectral SNR of the detection; (2) observational conditions reflecting the data quality such as the number of pointings covering the source, the array configurations of the pointings and the position of the source relative to the beam centre in each pointing.

Each evaluation criterion is assigned a binary number consisting of one to two bits. To be more precise, if we use a single bit for a criterion, a source can be divided into two categories, represented by boolean numbers ``1'' and ``0''. On the other hand, if we use two bits for a criterion, a source can be divided into four categories, represented by binary numbers ``11'', ``10'', ``01'', and ``00''. These binary numbers are used to represent data in a digital system, where each bit represents a binary digit (0 or 1) that is used to store and process information. After this classification process, we concatenate the binary representation of all criteria to form a new binary number. This newly formed binary number is then converted as a whole into its corresponding decimal equivalent, which will be explained in further detail with a specific criteria set and demonstrated through equation 2. Those binary numbers from the criteria placed as a priority (located among the first criteria considered) would contribute more than those of lower priority (which are among the last of the criteria to be considered). Ideally, a well-designed set of criteria (both in the sense of the number of possible classifications, 2 or 4, and in the order in which they are considered) would allow us to rank the sources properly from the most robust candidates to those with no evidence for extended molecular gas. 

\begin{figure*}
    \centering
    \includegraphics[width = 1.0\linewidth]{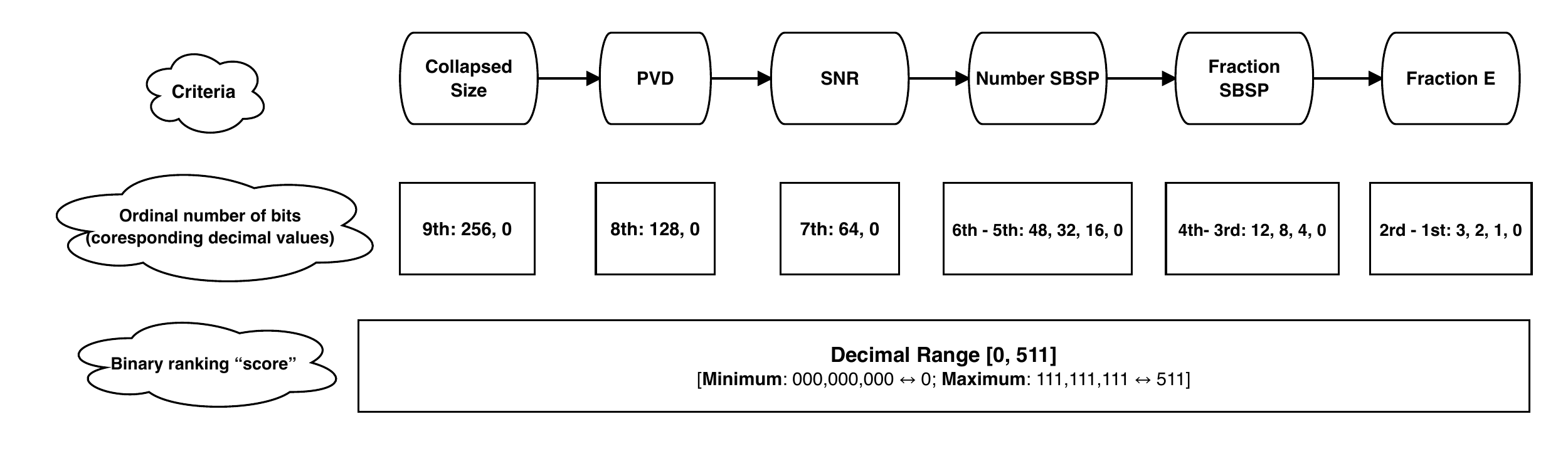}
    \caption{Display of the criteria order, and corresponding decimal values. The first row lists the criteria in order of their weighting, highest to lowest from left to right. The second row shows the corresponding decimal values each criteria could contribute. The third row is the overall corresponding decimal range of all possible combinations. A detailed presentation of quantifying these criteria into decimal values and classification of candidates is  shown in Figure~\ref{fig:CriteriaRanking__}.}
    \label{fig:Criteria_Ranking}
\end{figure*}

We designed the ranking system and adjusted the definition and order with seven ``calibrators'' of extended molecular gas reservoir candidates. Those seven calibrators are either confirmed to be extended with high-resolution observations or with evidence for having extended gas reservoirs in our visual inspection process. Five other criteria sets designed previously are explained in Appendix~\ref{sec:A0_CriteriaRankingDesign}. The criteria order and the corresponding decimal contribution of the final criteria set is shown in Figure~\ref{fig:Criteria_Ranking} (and a complementary Figure~\ref{fig:CriteriaRanking__} illustrates the detailed criteria used in assigning decimal values and in classifying galaxies as being a large gas reservoirs candidate). The binary to decimal conversion equation is,
\begin{equation}
\label{equ:Binary2Decimal}
Score = \sum_{i=0}^{N-1} x_i \cdot 2^i \quad\left(x_i=0 \text { or } 1\right)
\end{equation}
$x_i$ are the binary values at i-th places, and N is the total binary bits places for all the criteria. We will refer to the decimal values of the binary ranking as the "score", representing the possibility of being categorised as an extended molecular gas reservoir.

The six criteria, in order of priority, include the (1) collapsed image size (Collapsed Size); (2) features observed in the position-velocity diagram (PVD); (3) Signal-to-noise ratio(SNR); (4) The number of ATCA observational pointing covering the source with a small-beam-size ($<$7$\farcs$0) (Number of Small-Beam-Size Pointings - Number SBSP); (5) The fraction of small-beam-pointing pointings ($<$7$\farcs$0) relative to the total number of pointings that cover the source (Fraction of Small-Beam-Size Pointings - Fraction SBSP); (6) The fraction of pointings within which the source is located at the edge (Fraction E). The first three reflect the characteristics of each source, and the latter three the observational conditions. The detailed explanations of each criterion are given in Appendix~\ref{sec:A0_CriteriaRankingDesign}. Combine the binary numbers of each criteria and convert to decimal value, 

\begin{equation}
    \begin{split}
        &\text{[Collapsed size] ($x_8$)} \oplus \text{[PVD] ($x_7$)} \oplus \text{[SNR] ($x_6$)} \\
        & \qquad \oplus \text{[Number SBSP] ($x_5$ $x_4$)} \oplus \text{[Fraction SBSP] ($x_3$ $x_2$)} \\
        & \qquad \oplus \text {[Fraction E] ($x_1$ $x_0$)} \\
        & = \text{$x_8$ $x_7$ $x_6$ $x_5$ $x_4$ $x_3$ $x_2$ $x_1$ $x_0$} \quad \text{(combine)} \\
        & \text{$\Rightarrow$} \sum_{i=0}^{N-1} x_i \cdot 2^i (N=9) \quad \text{(convert to decimal)}
    \end{split}
\end{equation}

operator $\oplus$ means concatenate (combine) the binary numbers.

As shown in Figure~\ref{fig:Decimal_Distribution}, we separate the candidates from the rest via setting the cut-line (lower limit) on the corresponding decimal values 400. A detailed description about this chosen threshold is presented in Appendix~\ref{sec:DesignSeparationLine}.

\subsubsection{Calibration}
\label{sec:method_IV}
We use calibrators to improve the accuracy of our criteria by adjusting the number of bits and their order. Specifically, the adjustments we made are: (1) the order of the criteria representing the weights of each; (2) by adjusting the number of bits of one criterion, we can have different numbers of classification classes according to the specific criteria. One bit for two classes, two bits for four classes, three bits for up to eight classes, and so forth; and (3) adjusting the boundaries between sources with and without extended reservoirs.

The seven calibrators used to calibrate our classification method are as follows: Four of them are large gas reservoirs confirmed through the high-resolution data, the COALAS-SW.01 (HAE229: \citealt{Dannerbauer_2017AA...608A..48D}), the COALAS-SW.02 (Spiderweb Galaxy: \citealt{Emonts_2013MNRAS.430.3465E, Emonts_2018MNRAS.477L..60E}), the COALAS-SW.03 (DKB03: reported in this paper) and COALAS-SW.06 in Figure~\ref{fig:COALAS_06}. The remaining three have only the mosaic observations but they show clear indications of extended gas reservoirs: the COALAS-SW.29 which has an obvious merger/rotating-like kinematic, and the COALAS-SW.23 and COALAS-SW.46 are a close pair sharing a giant molecular reservoir. We give detailed information on COALAS-SW.06, COALAS-SW.29, COALAS-SW.23 and COALAS-SW.46 in Appendix~\ref{sec:A_CriteriaCalibrators}.

In total, we experimented with six different ranking sets with similar criteria that considered both source characteristics and observational conditions aspects. Given the importance of the collapsed size in searching for extended gas reservoirs, we assigned the highest priority to this criterion in five out of the six sets. During the implementation and adjustment of the criteria ranking sets, we realised the need to lower the weights of collapsed size criterion, to potentially allow other criteria to provide greater contribution to the final ranking. Specifically, according to the behaviours of the calibrators, the candidates could have sizes comparable to the beam size, and clear indications from the PVD analysis for an extended reservoir of gas. To lower the weight of collapsed size criterion, we reduced the bits numbers of collapsed size criterion from two to one. Furthermore, we lowered the standard to let in those not obviously observed as extended option V. In option VII, the PVD criterion was simplified to one bit, which gives more weight to the criteria besides collapsed size and PVD ones, and thus calibrators COALAS-SW.06 and COALAS-SW.46 become higher ranked.

Option VII in Table~\ref{tab:Criteria} is the final criteria set for ranking the sources (Figure~\ref{fig:Criteria_Ranking}) which works well in separating the promising candidates from the rest. The obvious break of decimal values around 400 seems to be the adequate criteria set (Figure~\ref{fig:Decimal_Distribution}). Improvements lead to a break in the ranking from the robust to the tentative one. A clear break implies that this final option is likely the best for producing a set of robust candidates of sources with extended gas reservoirs. Additionally, we include a final manual classification for separating the robust candidates and those which are rather tentative.

\subsection{Results}
\label{sec:sec2_results}
\subsubsection{Candidate List}
\label{sec:Result_Candidates}

\begin{figure}
    \centering
    \includegraphics[width = 1.0\linewidth]{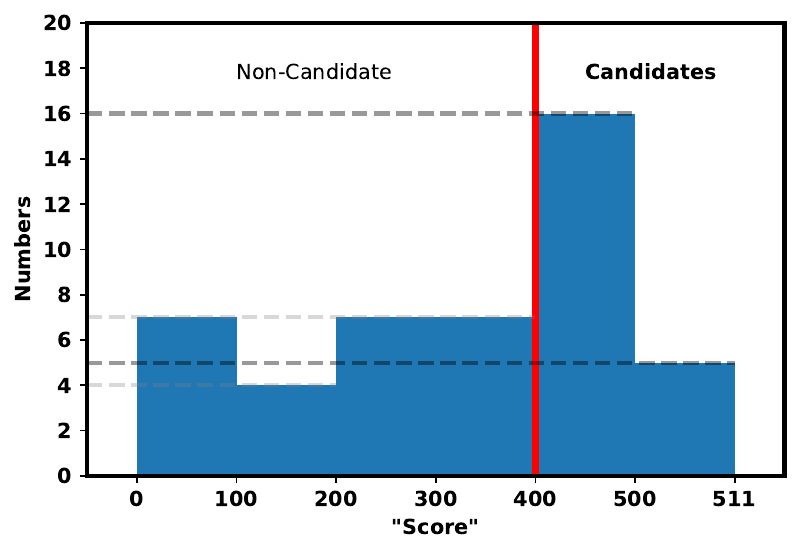}
    \caption{Binary ranking ``score'' distribution of COALAS sample. The red dashed line is the separation line for the large molecular gas reservoir candidates (robust and tentative) and the remaining ones.}
    \label{fig:Decimal_Distribution}
\end{figure}

As demonstrated above, the final set of criteria consists of six items, employing nine binary bits, and thus resulting in the corresponding decimal values ranging from 0 to 511.
\begin{equation}
\label{equ:maxmin}
\text { Score }=\sum_{i=0}^{N-1} x_i \cdot 2^i \stackrel{N=9}{===}\left\{\begin{array}{l}
0\;\;\;\;\;\left(\text {minimum, all } x_i=0\right) \\
511\;\left(\text { maximum, all } x_i=1\right)
\end{array}\right.
\end{equation}
The binary ranking ``score'' distribution of COALAS CO emitter sample is shown in Figure~\ref{fig:Decimal_Distribution}. The ``score'' value 400 is where the source number drops, and was used as a separation line (the red line in Figure~\ref{fig:Decimal_Distribution}) between candidates and non-candidates. The 21 sources whose scores are above 400 are our final candidates.

In accordance with our classification system outlined in Stage V of Section~\ref{sec:method_I}, all identified extended molecular gas reservoir candidates were further categorised into two distinct groups: robust and tentative candidates. Collapsed images and PVDs for the 14 robust and seven tentative candidates are presented in Appendix~\ref{sec:A1_Robust} and \ref{sec:A2_Tentative}, respectively. The frequency of extended gas reservoirs in the Spiderweb Protocluster is $\sim$30 $\%$ (14 out of 46 galaxies), counting the robust candidates only.

\subsubsection{Environmental study}
\label{sec:Result_Env}
Due to the handful of discoveries and studies on extended molecular gas reservoirs, their nature is still unclear. For massive BCGs, the large gas reservoirs were suggested to be either molecular gas in accretion streams or recycled gas from the star formation within the galaxy~\citep{Emonts_2016Sci...354.1128E, Emonts_2018MNRAS.477L..60E, Ginolfi_2017MNRAS.468.3468G}. However, our extended gas reservoir candidates were discovered to be spatially distributed over the Spiderweb protocluster field, not limited to the central galaxies like BCGs. Thus, we cannot directly apply the aforementioned scenarios to these candidates and require new explanations for their nature. Nevertheless, if these scenarios are generalised, the denser local potential well in the nodes of the cosmic web may have a similar impact on gas accretion and interactions between close-by galaxies as the (proto)cluster centre.

The measurement of the galaxy environment is carried out in three ways: N-th nearest neighbour, fixed aperture and annulus (refer to Table 1 in \citealt{2012MNRAS.419.2670M} for a review). We use the N-th nearest neighbour method to determine galaxy density by counting the number and proximity of neighbouring galaxies. Based on such a simple concept, one can either define a projected surface density or a spherical enclosed density. The projected surface density, ${\sigma}_n$, is defined as 
\begin{equation}
\sigma_{n}=\frac{n+1}{\pi r_{n}^{2}}
\end{equation}
where n is the N-th closest galaxy for each reference galaxy, and $r_n$ is the projected distance between the reference galaxy and the N-th closest galaxy. This is commonly used for datasets lacking the third space dimension, i.e., precise redshift (e.g., \citealt{1980ApJ...236..351D}, \citealt{2006MNRAS.373..469B}, \citealt{Koyama_2013MNRAS.434..423K}). A severe shortcoming of the 2D surface density is that two galaxies appearing close to each other could be due to pure alignment along the line-of-sight and being separated at large distances in the third dimension. Among the 46 CO emitters in the COALAS survey, several projected close-by sources have large distances between each other in the three-dimensional aspect. Therefore, we adopt the 3D-type N-th nearest neighbour method.

\begin{figure}
    \centering
    \includegraphics[width = 1.0\linewidth]{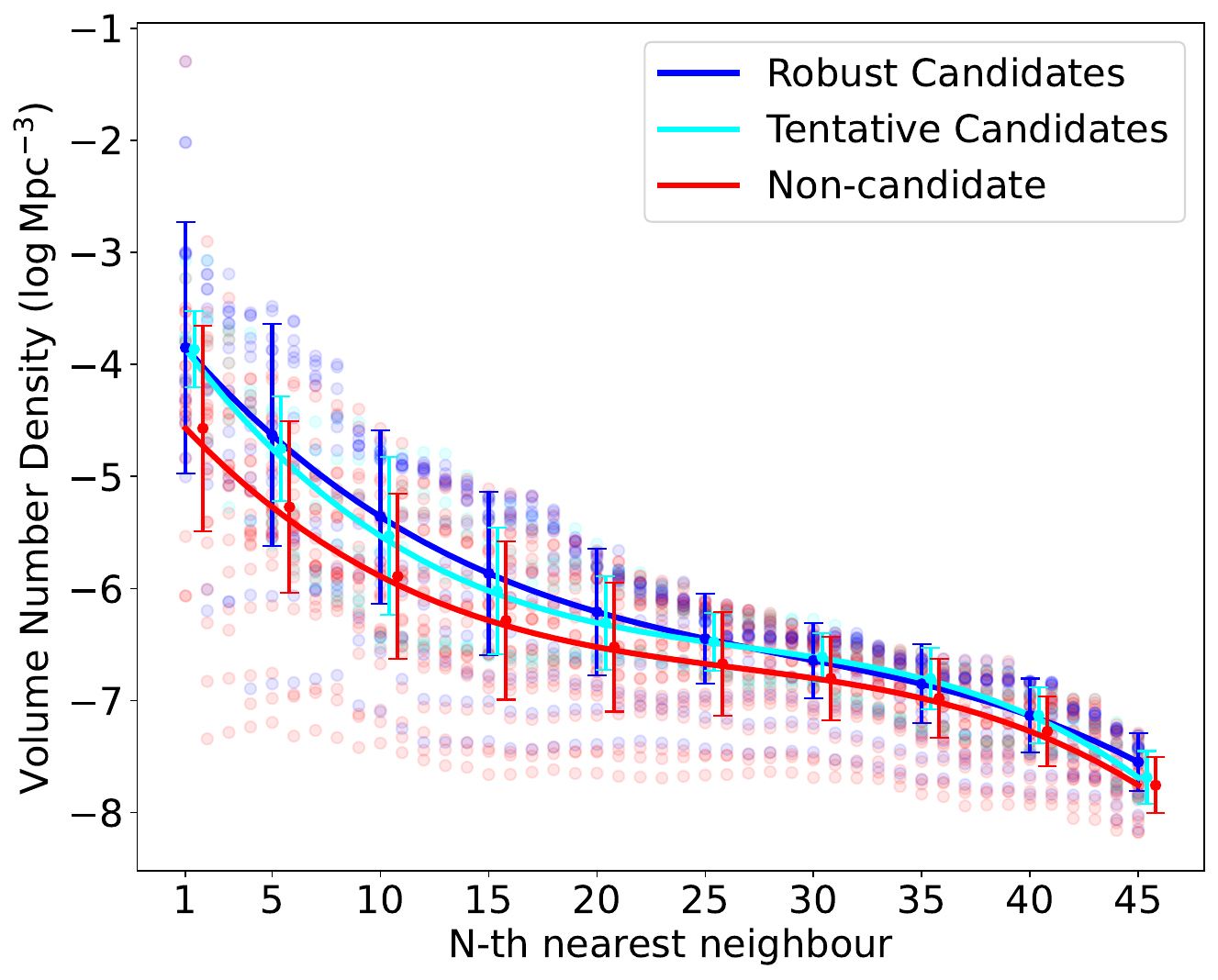}
    \caption{Volume density to N-th nearest neighbour of CO emitters. Robust candidates, tentative candidates and the non-candidates were marked with blue, cyan and red colours, with solid lines indicating best fit polynomials. Error bars for the fitting are shown at intervals of every 5th nearest neighbour, with specific horizontal adjustments applied to the error bars of the robust, tentative, and non-candidates classes to enhance the visual presentation.}
    \label{fig:Result_Env}
\end{figure}

Utilising the precise 3-dimensional location information available for all 46 sources, we conducted a calculation of the 3D spherical N-th nearest neighbour for each individual source. The corresponding volume density of galaxies is defined in the following way:
\begin{equation}
\rho_{n}=\frac{n+1}{(4 / 3) \pi r_{n}^{3}}
\end{equation}
where n is the N-th closest galaxy of each reference galaxy and $r_n$ is the distance between the reference galaxy and the N-th nearest galaxy in the three-dimensional space. Our environmental analysis focuses on the 46 CO detections ($\geq$ 3.8) reported in \citet{Jin_2021AA...652A..11J}, utilising the corresponding coordinates and CO spectroscopic redshift data (precise to four decimal places) from the same study.
In our sample, each galaxy has 45 neighbours. Regarding each of them as the centre reference galaxy, we calculated the volume density from 1st to 45th nearest neighbour. We plot the volume density to N-th nearest neighbour in Figure~\ref{fig:Result_Env}. We separated the robust candidates, tentative candidates and the rest with blue, cyan, and red colours, fitted with binomial lines. We found that our candidates tend to be located in denser regions.  We note that the contrast on density between candidates and non-candidates remains even when fitting the non-candidates without the four sources located at the sparsest environment.

The interactive Figure~\ref{fig:FinalCandidates_Distribution3D} displays the 3D distribution of CO emitters. The robust, tentative and non-candidates are marked with blue squares, cyan dots and red crosses, respectively. The Spiderweb galaxy is marked with a blue star. Alternatively, Figure~\ref{fig:Source_Distribution} offers a two-panel view of the spatial and redshift distributions of 46 CO emitters. The upper panel displays their spatial distribution, categorised into non-candidates (red crosses), robust candidates (blue circles), and tentative candidates (cyan circles), with grey crosses indicating the distribution of all CO emitters. The lower panel illustrates their redshift distribution, with Gaussian functions used to fit the distributions for each category, with the fitting lines denoted in the same colours as their respective categories.

We find that the large gas reservoir candidates are concentrated in the core of the Spiderweb protocluster, which is consistent with the results of the N-th nearest neighbour analysis. Spatially, as shown in Figure~\ref{fig:FinalCandidates_Distribution3D} or the upper panel of Figure~\ref{fig:Source_Distribution}, the central region is dominated by candidates, while the outskirts are conversely occupied more by non-candidates. In terms of redshift, in contrast to the almost uniform distribution of non-candidates, the candidates are highly concentrated at a redshift of approximately z=2.16, which is the centre of the Spiderweb Protocluster.

The inhomogeneous mosaic data have different local RMS levels varying by up to a factor 2 (0.13 - 0.29 mJy) among the 13 pointings, e.g., pointing ``MRC1138'' and ``HAE229'' are the two deepest fields with rms of value 0.13 mJy, and ``SWpoint7'' has the largest rms of value 0.29 mJy (refer to the Table 1 in \citealt{Jin_2021AA...652A..11J}). This raises the question if the aforementioned result, extended gas reservoirs tend to be located at the denser region, is caused by an observational bias. We thus checked if there is a correlation between the location of large gas reservoir candidates and local RMS, and found no evidence of such a correlation.

\begin{figure}
    \centering
    \animategraphics[controls,loop,width=1.0\linewidth]{2}{./Animation_3D/SD_}{1}{36}
    \caption{Spatial distribution (3D) of all 46 CO emitters. Blue star: Spiderweb galaxy; blue square: robust candidates; cyan: tentative candidates; red: the others.}
    \label{fig:FinalCandidates_Distribution3D}
\end{figure}

\begin{figure}
    \centering
    \includegraphics[width = 1.0\linewidth]{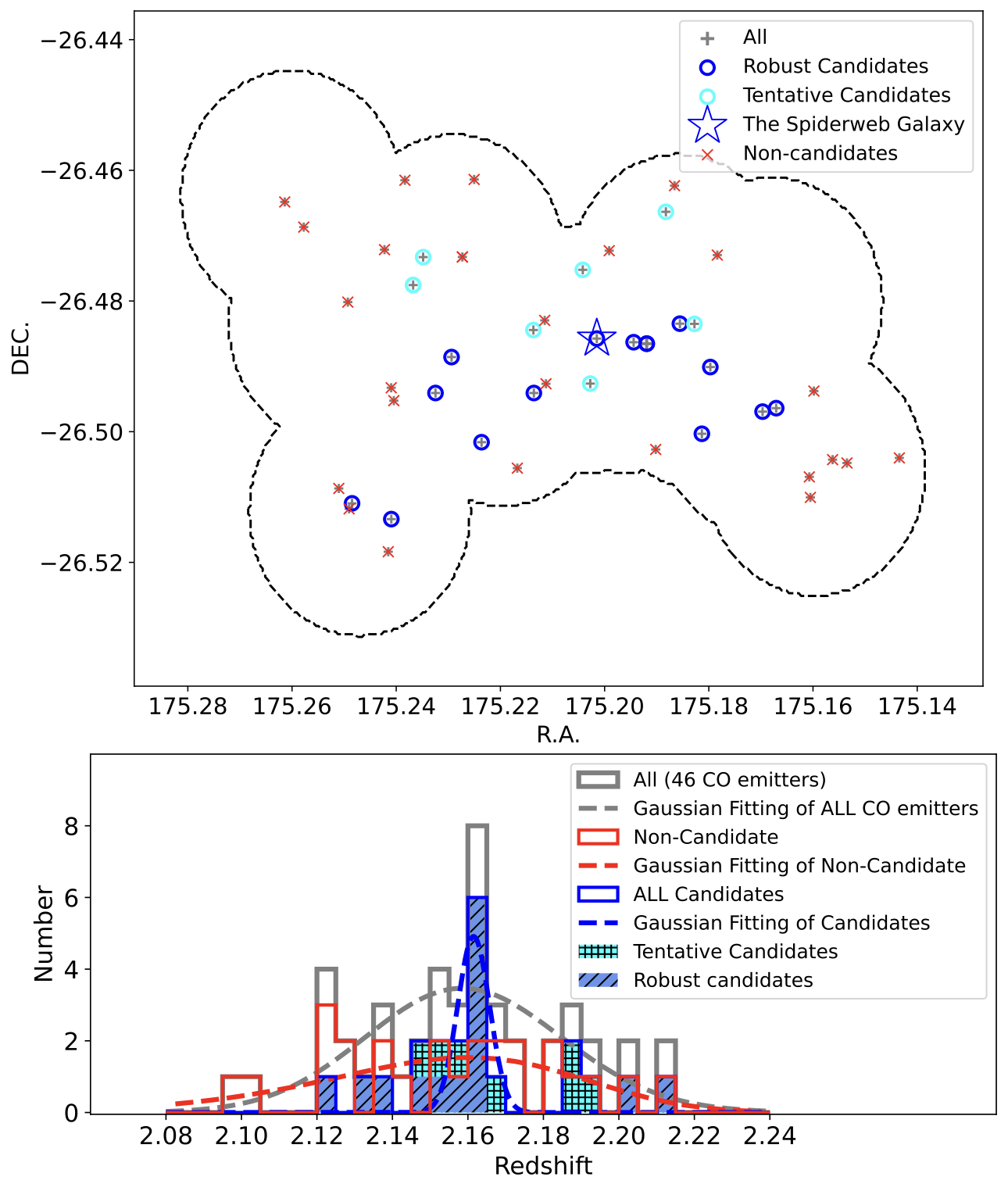}
    \caption{The upper panel displays the spatial distribution of 46 CO emitters, including non-candidates (red crosses), robust candidates (blue circles), and tentative candidates (cyan circles), with grey crosses indicating the distribution of all CO emitters. The Spiderweb Galaxy is denoted by a blue star. The lower panel illustrates the distribution of redshifts among the CO emitters, divided into non-candidates (in red) and candidates (in dark blue), with the overall CO emitter distribution depicted in grey. The distribution for candidates is further broken down into two categories: tentative candidates (represented by the cyan shaded region) and robust candidates (represented by the blue shaded region). Gaussian functions have been employed to fit the distributions, with corresponding fitting lines in the same colours as the respective categories.}
    \label{fig:Source_Distribution}
\end{figure}

\section{Discussion}
\label{sec:discussion}
\subsection{Performance of our method}
We obtained 14 robust candidates of large molecular gas reservoirs (i.e., $\sim$30$\%$ of the total number of candidates). The following aspects support the credibility of our method and final robust candidate list: (1) The binary criteria ranking system was calibrated with seven CO detections (four of the seven calibrators are cases confirmed through high-resolution observations). A clear border between the candidates and non-candidates resulted in the number distribution of decimal ranking values (Figure~\ref{fig:Decimal_Distribution}). We note that follow-up high-resolution observations of our candidates are needed for a verification. (2) Our environmental study of the candidates is in line with the physical expectations that the gas is accumulating and collapsing in the denser regions (especially in the cores of protoclusters) where the potential well is deeper.

\subsection{Large molecular gas reservoirs from the Literature}
We conducted a comprehensive search of molecular gas observations of (proto)clusters and overdensities aiming to identify potential extended gas reservoirs with a scale of $\gtrsim$40 kpc (corresponding to the scale of large gas reservoirs we aimed to identify in the Spiderweb protocluster in the current study). The compilation is listed in Table~\ref{tab:collection} and focuses on (multiple) transition(s) of CO. One source, LAB1 in the SSA22 protocluster field, was revealed through [CII] observations. In total we have identified 13 large molecular gas reservoirs. Four sources have been already reported before as such objects: the Spiderweb galaxy~\citep{Emonts_2013MNRAS.430.3465E,Emonts_2014MNRAS.438.2898E,Emonts_2016Sci...354.1128E,Emonts_2018MNRAS.477L..60E}, HAE229~\citep{Dannerbauer_2017AA...608A..48D} in the Spiderweb protocluster, the QSO Q12287+3128 in the ELANe protocluster survey~\citet{LiJianrui_2021ApJ...922L..29L,Li_2023arXiv230402041L} and MAMMOTH-I in the BOSS1441 protocluster \citep{Emonts_2018MNRAS.477L..60E}. The size scales of the remaining sources were determined through visual measurements on moment 0 maps based on their contours at approximately 3$\sigma$. However, our exercise could be only done in cases where these maps have been provided. Thus our literature search might be not complete.

\begin{table*}
    \centering
    \begin{tabular}{c|c|c|c|c|c|c}
        Protocluster & redshift & Source ID & Emission Lines & Size & Size Given & Reference \\
        \hline\hline
        Jackpot nebular & 2.04 & galaxy1 & CO(3-2) & $\sim$40 kpc & -  & \citealt{Decarli_2021AA...645L...3D} \\
        \hline
        PKS1138-262 Protocluster & 2.16 & Spiderweb Galaxy & CO(1-0) & $\sim$70 kpc & Yes & \citet{Emonts_2013MNRAS.430.3465E, Emonts_2014MNRAS.438.2898E, Emonts_2016Sci...354.1128E, Emonts_2018MNRAS.477L..60E}\\
        (Spiderweb) & & & CO(4-3) & $\sim$50 kpc & Yes & \citet{Emonts_2018MNRAS.477L..60E} \\
         & & & [CI] & $\sim$50 kpc & Yes & \citet{Emonts_2018MNRAS.477L..60E}\\
         \cline{3-7}
         & & HAE229 & CO(1-0) & $\sim$40 kpc & Yes & \citet{Dannerbauer_2017AA...608A..48D} \\
        \hline
        Protocluster ELANe & 2.22 & QSO  Q12287+3128 & CO(4-3) & $\sim$100 kpc & Yes & \citet{LiJianrui_2021ApJ...922L..29L,Li_2023arXiv230402041L} \\
        \hline
        Slug nebular & 2.28 & QSO & CO(3-2) & $\sim$50 kpc & - & \citet{Decarli_2021AA...645L...3D} \\
        \hline
        BOSS1441 Protocluster & 2.3 & Region A (Q0052) & CO(1-0) & $\sim$40 kpc & - & \citet{Emonts_2019ApJ...887...86E} \\
        (MAMMOTH-I) & & & CO(3-2) & $\lesssim$15 kpc & - & \citet{Li_Qiong_2021ApJ...922..236L}\\
         & & & CO(4-3) & $\lesssim$15 kpc & - & \citet{Li_2023arXiv230402041L} \\
        \hline
        CLJ1001 Protocluster & 2.5 & 131077 & CO(1-0) & $\lesssim$40 kpc & Yes & \citet{Champagne_2021ApJ...913..110C} \\
         & & & CO(3-2) & $\lesssim$10 kpc & - & \citet{Champagne_2021ApJ...913..110C}\\
         & & & CO(1-0) & $\sim$30 kpc & - & \citet{Wang_2016ApJ...828...56W}\\
         & & & CO(5-4) & $\sim$30 kpc & - & \citet{Wang_2016ApJ...828...56W}\\
         \cline{3-7}
         & & 130933 & CO(1-0) & $\sim$60 kpc & - & \citet{Wang_2018ApJ...867L..29W} \\
         \cline{3-7}
         & & 130842 & CO(1-0) & $\sim$60 kpc & - & \citet{Wang_2018ApJ...867L..29W} \\
        \hline
        HXMM20 Protocluster & 2.6 & S0 & CO(1-0) & $\sim$45 kpc & - & \citet{Gomez-Guijarro_2019ApJ...872..117G}\\
         & & & CO(3-2) & $\sim$30 kpc & - & \citet{Gomez-Guijarro_2019ApJ...872..117G}\\
         \cline{3-7}
         & & S2 & CO(1-0) & $\sim$40 kpc & - & \citet{Gomez-Guijarro_2019ApJ...872..117G}\\
         & & & CO(3-2) & $\sim$30 kpc & - & \citet{Gomez-Guijarro_2019ApJ...872..117G}\\
         \cline{3-7}
         & & S3 & CO(1-0) & $\sim$40 kpc & - & \citet{Gomez-Guijarro_2019ApJ...872..117G}\\
         & & & CO(3-2) & $\sim$30 kpc & - & \citet{Gomez-Guijarro_2019ApJ...872..117G}\\
        \hline
        SSA22 Protocluster & 3.1 & LAB1 & [CII] & $\sim$50 kpc & - & \citet{Umehata_2017ApJ...834L..16U,Umehata_2021ApJ...918...69U}\\
         & & & CO(4-3) & $\sim$30 kpc & - & \citet{Umehata_2021ApJ...918...69U}\\
        \hline\hline
    \end{tabular}
    \caption{Collection from the literature of potential extended molecular gas reservoirs with scale $\gtrsim$ 40 kpc. In addition to emission lines indicative of a scale above 40 kpc, other relevant emission lines of the same sources are also listed exhibiting compact or less extended behavior.}
    \label{tab:collection}
\end{table*}

\begin{figure}
    \centering
    \includegraphics[width=1.0\linewidth]{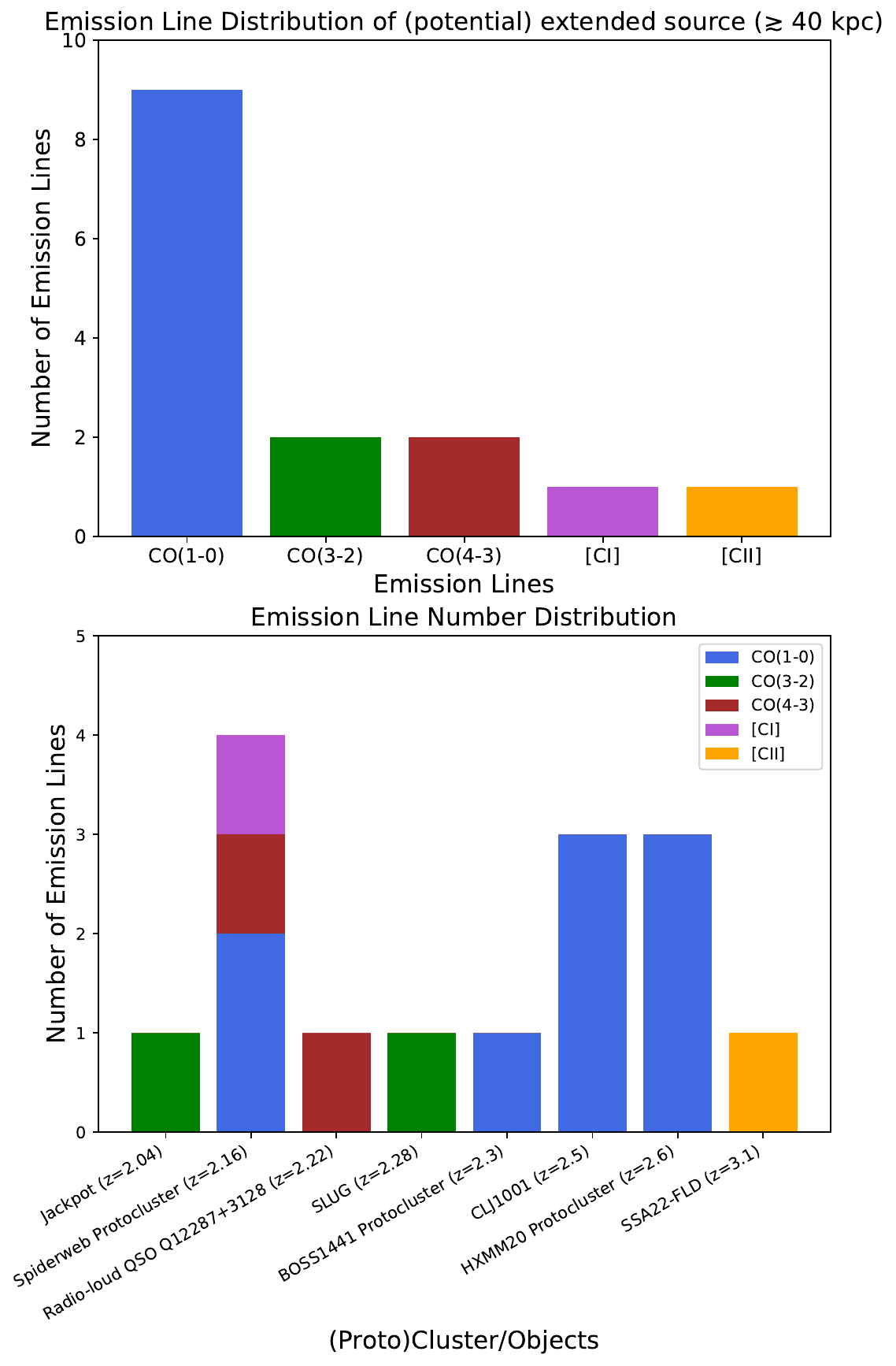}
    \caption{The upper panel displays the distribution of emission line numbers for (potential) extended gas reservoirs, while the bottom panel illustrates the situation of each individual (proto)cluster/object.}
    \label{fig:PotentialExtended_Literature}
\end{figure}

We note that some protoclusters lack molecular gas observations while others only had relevant observations for the central regions or focused solely on the central BCGs. The absence of COALAS-like wide field systematic gas surveys in these protoclusters may have contributed to the rarity of extended gas reservoir discoveries reported up to now in the literature. For six cases, we have several CO transitions. Only in the case of the Spiderweb Galaxy all reported transitions show extended emissions. In the remaining cases the phenomena is only seen in the lowest CO transition CO(1-0). Statistically, about half of our (potential) large gas reservoirs are selected based on CO(1-0) observations, as shown in the upper panel of Figure~\ref{fig:PotentialExtended_Literature}. Increasing the number CO(1-0) observations of (proto)cluster members would certainly lead to more detections of extended gas reservoirs. Recent studies show that the extended molecular gas reservoirs are not only limited to AGNs located in protoclusters, e.g., star-forming galaxy HAE229 in Spiderweb protocluster~\citep{Dannerbauer_2017AA...608A..48D}, and the COALAS-SW.03 reported in this paper (starburst galaxy dubbed DKB03 in ~\citealt{Dannerbauer_2014AA...570A..55D}). 

Existing observations in the literature may have downplayed potential large gas reservoirs, and some sources of large gas reservoirs may have been missed due to inadequate observations. Specifically, the low-surface-brightness ground-transition of CO has not been properly observed, and the lack of systematic observations has resulted in some sources being completely overlooked as based on our visual inspection we reported 13 new extended molecular gas reservoirs. With our work at least we doubled the number of known extended molecular gas reservoirs.

\subsection{Environmental-driven physical process: thirty percent of galaxy members are evolving in extended gas}
With our method, we find 14 robust candidates of extended molecular reservoirs selected from 46 CO emitters physically related to the Spiderweb protocluster. Thirty percent is a previously unknown large fraction, which means extended gas might be prevalent in protoclusters and, due to the difficulty in detecting extended low surface brightness gas, was generally not properly characterized in the past. More observations and theoretical studies are needed to understand this result.

Converting the CO line luminosity to the molecular gas mass, we can compare the ICM in the Spiderweb protocluster with local galaxy clusters (e.g., Coma and Virgo clusters) and conjecture the future of the Spiderweb protocluster. A bimodal CO-$H_2$ conversion factor is widely adopted in the extragalactic literature. For star-forming galaxies, commonly the conversion factor \(\alpha=4.6 \mathrm{M}_{\odot}\left(\mathrm{Kkms}^{-1} \mathrm{pc}^2\right)^{-1}\) is thought to be most appropriate. However, for galaxies with intense star formation or mergers, which bear resemblance to local ultra-luminous infrared galaxies, \(\alpha=0.8 \mathrm{M}_{\odot}\left(\mathrm{Kkms}^{-1} \mathrm{pc}^2\right)^{-1}\) is adopted by many researchers who investigate high-redshift CO emitters~\citep{Solomon_2005ARAA..43..677S, Casey_2014PhR...541...45C}. Assuming a single galaxy population, we obtained the molecular gas mass for each of the 46 CO emitters, and the total molecular mass is 1.5 $\times$ ${10}^{13}$ (starburst galaxies) - 8.7 $\times$ ${10}^{13}$ $M_{\odot}$ (star-forming galaxies). For the molecular gas in extended morphologies (14 out of 46 robust candidates; thirty percent of the full sample), the total molecular gas mass is 3.5 $\times$ ${10}^{12}$ (starburst galaxies) - 2.0 $\times$ ${10}^{12}$ $M_{\odot}$ (star-forming galaxies), and this accounts for 23$\%$ of the total molecular gas mass. Taking all the extended molecular gas and assuming the low molecular-to-atomic ratio appropriate for local galaxies ($\sim$0.1; \citealt{Saintonge_2017ApJS..233...22S, Catinella_2018MNRAS.476..875C}), we got a total $\sim$ ${10}^{13}$ - ${10}^{14}$ $M_{\odot}$ of gas for just the galaxies with extended reservoirs of gas. The mass of Virgo cluster derived from hot gas is $\sim$ 3 $\times$ ${10}^{14}$ within the central 1.7 Mpc~\citep{sparke_gallagher_iii_2007}. The Spiderweb protocluster has sufficient amount of gas to evolve into a Virgo-like cluster, and through the ``truncation'' process \footnote{galaxy-galaxy encounter or gravitational interactions between galaxies and the (proto)cluster environment can result in the distortion, stripping, and truncation of galaxy halos \citealt{Moore_1996Natur.379..613M, Moore_1998ApJ...495..139M, Fujita_1998ApJ...509..587F}.}. With these assumptions, it is plausible that if stripped of their gas, these large gas reservoirs may contribute the cold neutral and molecular phases of a proto-intracluster medium. We note the diverse dependencies of CO-$H_2$ conversion factors, including metallicity, gas density, temperature, and radiation field~\citep{Pavesi_2018ApJ...861...43P, Madden_2020AA...643A.141M}. Estimations using bimodal conversion factors are simplified. Therefore, follow-up observations and detailed studies are crucial for better constraining the gas content, considering these various factors.

Based on the environmental study carried out in Section~\ref{sec:Result_Env}, it is found that the large molecular gas reservoirs candidates are located in a locally denser environment, predominantly concentrated in the central region of the Spiderweb protocluster. According to~\citet{Jin_2021AA...652A..11J}, the Spiderweb protocluster is likely a super-protocluster that is embedded within a larger-scale filamentary structure, and perhaps half of the CO emitters are still not bounded to the core region. Hence, it is possible that the Spiderweb protocluster comprises numerous substructures, and that the constituent galaxies evolved collectively within separate relatively massive sub-halos. Our findings suggest that the concentration of galaxies with extended reservoirs of molecular gas in denser regions of protoclusters may be indicative of efficient accretion of cold gas flowing along filaments, which could potentially fuel the growth of the stellar population within these galaxies. This could explain the observed concentration of massive early type galaxies in local clusters. 

However, the origin of these extended reservoirs is unclear from available data. If the concentration of galaxies indeed reflects the overall and/or local gravitational potential, then it follows that these galaxies are positioned near the (local) centre(s) of the potential well. Consequently, one can expect the accretion rate of gas to be focused in these regions. Of course, whether or not gas accretion continues in massive halos and redshifts like those of the Spiderweb protocluster is not clear~\citep{Dekel_Birnboim_2006MNRAS.368....2D,Cornuault_2018AA...610A..75C}. Moreover, cold streams could potentially penetrate the early, massive, and hot halos and serve as the primary mode of galaxy formation~\citep{Dekel_2009Natur.457..451D}. Nevertheless, theoretically it is clear that gas accretion should create large extended reservoirs of gas (e.g., \citealt{Danovich_2015MNRAS.449.2087D}).  We can demonstrate what we mean by this through a few simple timescale estimates. For example, let us assume that these extended reservoirs of gas have order motions.  We can calculate a dynamical time, based on the assumption that the angular momentum resembles that of a rotating-like system, yields an approximate value of t$_{orbital}\sim$1.2 Gyr (r/40 kpc)/(v$_{orb}$/200 km s$^{-1}$). The gas cannot inspiral into the galaxy potential, say through tidal effects due to bars or spiral arms or other disk substructure in less than a dynamical time.  Such a long timescale, although crudely estimated, suggest that these extended reservoirs of molecular gas could be long lived if not disrupted by external forces (e.g., ram pressure stripping, tidal stipping by passing/merging gas, etc). Furthermore, if the extended gas is stable, it can fuel the future growth of these proto-cluster galaxies for a long time, at least 1 Gyr. So even if gas accretion has ceased, as galaxies fall into the proto-cluster potential, there is still enough remaining extended gas to support their growth and star formation rates for approximately the same time scale. 

Even more curiously, we can again estimate a crude crossing timescale if indeed the Spiderweb proto-cluster is forming and the galaxies are falling into the potential. Again, just making simple assumptions to provide a sense of the timescales we are discussing, we can estimate the crossing time or infall time, albeit crudely, as t$_{crossing}\sim$1 Gyr (r/Mpc)/(v$_{infall}$/1000 km s$^{-1}$) \citep{Kuiper_2011MNRAS.417.1088K}. This is the timescale over which processes like tidal and ram pressure stripping may play a role in truncating the gas distribution in these galaxies. Comparing the two crude estimates of the dynamical time of the extended reservoirs of gas and the crossing or infall time of a galaxy, they are approximately equal (order-of-magnitude only). Even with this crude comparison, it begs the question, why do we observe these at all \citep{Dannerbauer_2017AA...608A..48D}? Observing such reservoirs is even more puzzling given the recent detection of the S-Z effect in the SpiderWeb protocluster \citep{Di_Mascolo_2023Natur.615..809D}. 

Unfortunately, our results raise more questions than they answer. What we have shown is that extended gas reservoirs may be common and their mere existence pose many interesting problems and questions. Perhaps these reservoirs are one of the smoking guns of gas accretion.  Smoking gun may be particular appropriate as a metaphor as these extended reservoirs may be the residual after accretion has ceased.  We simply do not know. If the extended molecular gas is long-lived, then why was the gas removed by tidal stripping and ram pressure stripping despite this being a proto-cluster and there is evidence for a (proto-)intracluster medium. Perhaps the galaxies without extended reservoirs are the galaxies that were stripped of their material, they may have contributed to the gas in the ICM. However, it may be that the extended gas is relatively metal poor, but the ICM of local clusters is metal-rich, generally containing more metals than the galaxies in the clusters. Therefore, there must be an ongoing exchange of material between the galaxies and their surrounding large molecular gas reservoirs, which is due to outflow generated by young massive stars, type Ia SNe and AGN through their radiation pressure and radio jets~\citep{Tumlinson_2017ARAA..55..389T}. This is evidenced by the elevated levels of gas metallicity observed in the Spiderweb Galaxy and its halo gas and HAE229~\citep{Emonts_2016Sci...354.1128E, Dannerbauer_2017AA...608A..48D}. In other words, what ever processes are dominating the nature of the gas in the centre of the protocluster, it is more complex than the simple picture of gas accretion of relatively pristine gas and outflows of metal-rich gas via starburst- and AGN-drive outflows. If the most simple picture (caricature) of inflows and outflow were correct, the extended gas reservoirs would become diluted due to the continuous accretion of relatively pristine gas via filaments. This scenario would lead to a result that is opposite to that observed in the Spiderweb Galaxy and HAE229. Other processes such a mixing and gas instabilities must allow inflows and outflows to enrich or dilute each other. If not, these extended reservoirs become virtually impossible to explain with any reasonable mass.

Furthermore, it was first discovered in \citet{Hatch_2008MNRAS.383..931H} that the Spiderweb Galaxy had an extended stellar halo, which was postulated to originate from intense star formation fueled by a large (in size) gas reservoir. It was later, in 2016, that \citet{Emonts_2016Sci...354.1128E} discovered this gas reservoir, which nicely closed the loop on this picture. The gas reservoir appears to be forming stars at a rate consistent with the Kennicutt-Schmidt law. These stars are likely to be an early source of intracluster light, further adding to the complexity of the processes at play.

\section{Summary}
We present a method for searching for large molecular gas reservoirs in 46 CO emitters from \citet{Jin_2021AA...652A..11J} physically related to the $z=2.16$ Spiderweb protocluster. To show the feasibility of our method on extended source searching based on mosaic data, we took COALAS-SW.03 as an example, and compared the kinematic and morphological features obtained from the high-resolution and mosaic data. We classify CO emitters through visual inspections of collapsed images, PVDs, channel maps and integrated spectra. We propose a binary criteria ranking approach to quantify whether a source is extended or not. The criteria for ranking reflects both the source characteristics and the observational conditions of each source. The method is calibrated by seven sources, including four large gas reservoirs confirmed with high-resolution observations, and three having strong evidences of extended gas reservoirs solely based on mosaic observations. The major results we obtained in this work are:
\begin{itemize}
    \item We find 14 robust and seven tentative extended gas reservoir candidates and present their collapsed images and PVDs in Appendix~\ref{sec:A1_Robust} and \ref{sec:A2_Tentative}. The rate of cluster members containing large gas reservoirs is $\sim$30$\%$, and up to $\sim$50$\%$ if including the tentative detections.
    \item We collected 13 potential extended gas reservoirs in dense environments from literature, and highlighted the inadequacy of ground-transition CO observations, which has resulted in limited discoveries of large gas reservoirs.
    \item Through the analysis of the three-dimensional distribution we show that the candidates are concentrated on the core of the Spiderweb protocluster. Through our environmental study employing the N-th nearest neighbour method, we find that the candidates tend to be located in the local denser environments. This is in line with the scenario that gas was accreted efficiently by gravity in the denser region where the potential well is deeper.
    \item We discuss the underlying environmental-driven physical processes of the large molecular gas phenomenon. The large gas reservoirs might involve mechanisms like gas truncation, galaxy merger or interaction in a dense protocluster, and may feed the ICM in future local galaxy clusters like Virgo or Coma.
\end{itemize}

\section*{Acknowledgements}
We thank the anonymous referee for her or his comments that helped us to improve our arguements and presentation in this paper.
ZC acknowledges the support by China Scholarship Council (CSC). 
The National Radio Astronomy Observatory is a facility of the National Science Foundation operated under cooperative agreement by Associated Universities, Inc. The Australia Telescope is funded by the Commonwealth of Australia for operation as a National Facility managed by CSIRO. 
HD and JMRE acknowledge financial support from the Agencia Estatal de Investigación del Ministerio de Ciencia e Innovación (AEI-MCINN) under grant (La evolución de los cíumulos de galaxias desde el amanecer hasta el mediodía cósmico) with reference (PID2019-105776GB-I00/DOI:10.13039/501100011033). HD acknowledges support from the ACIISI, Consejería de Economía, Conocimiento y Empleo del Gobierno de Canarias and the European Regional Development Fund (ERDF) under grant with reference PROID2020010107. QSGU acknowledges support from the National Natural Science Foundation of China (No. 12192222, 12192220 and 12121003). TK acknowledges financial support by JSPS Kakenhi ($\#$18H03717) and International Leading Research ($\#$22K21349).

\section*{Data Availability}
This work is primarily based on the ATCA large program COALAS survey (ID: C3181, PI: H. Dannerbauer). The other relevant ATCA observations used in this work includes 2014OCTS/C3003 (PI: H. Dannerbauer), 2016APRS/C3003 (PI: H. Dannerbauer) and 2017APRS/C3003 (PI: B. Emonts). 

\bibliographystyle{mnras}
\bibliography{A.Paper} 

\appendix
\section{Criteria Ranking Design and Decimal Separation Line}
\subsection{Criteria Ranking Design}
\label{sec:A0_CriteriaRankingDesign}
To establish a robust set of criteria for accurately assessing the likelihood of source extension, we thoroughly identified relevant factors and defined their parameters. These parameters were then encoded into binary values consisting of one or two bits. By concatenating these binary values and converting them into decimal representation, we ranked the sources accordingly. Throughout the process, we experimented with various sets of criteria and ultimately selected option VII as the final decision. In the subsequent paragraphs, we present the details of option VII and elucidate the adjustments made to arrive at the final version.

In accordance with the details provided in Section 4.1, a set of seven calibrators, known either for their confirmed extension or for exhibiting evidence of extended gas reservoirs during our visual inspection process, was employed to assess and refine the accuracy of our method, specifically the criteria sets.

In the initial set of criteria, denoted as Option I, we included five key factors for assessment: (1) collapsed image size (Collapsed Size I), (2) features observed in the position-velocity diagram (PVD I), (3) the number of ATCA observational pointings covering the source with a small beam size (<7") (Number of Small Beam Size Pointings - SBSP), (4) the fraction of small-beam-size pointings (<7") relative to the total number of pointings that cover the source (Fraction of Small Beam Size Pointings - Fraction SBSP), and (5) the fraction of pointings within which the source located at the edge (Fraction E). We allocated two bits to each criterion, enabling the classification of a source into one of four distinct classes for each criterion.

Building upon the initial set of criteria, our aim was to enhance its effectiveness and accuracy by evaluating the ranking and inclusion of our seven calibrators in the final candidate list of extended molecular gas reservoirs. In Option II, our objective was to integrate the signal-to-noise ratio (SNR) of CO detections from Jin et al. (2021) into the existing criteria framework (Option I). In accordance with the SNR = 5 threshold specified in Table 3 of Jin et al. (2021), we assigned a single bit number to this criterion. Sources with an SNR greater than 5 were assigned a value of 1, while the remaining sources received a value of 0. Subsequently, in Option III, we explored a rearrangement of the first and second criteria compared to the initial set (Option I). Option IV involved combining Option I and Option II, maintaining the original order of the first two criteria, with emphasis on the "Size" criterion as our primary concern in selecting large molecular gas reservoirs. Additionally, we introduced the SNR criterion after Collapsed Size and PVD I criteria to further refine our selection process. Upon examining the outcome of Option IV, which considered the ranking and inclusion of our seven calibrators, and a thorough analysis of source properties, we identified that the Collapse Size criterion was over-weighted. To address this, adjustments were made in Option V, involving a reduction in the bit number from two to one and a redefinition of the collapsed size, leading to a refined classification of each source's Collapse Size properties into two distinct classes. (Collapsed Size II) \footnote{In the \underline{Collapsed Size I} classification, we assign "11" to sources with collapsing contours that are notably large compared to the pointing beam size and exhibit interesting shapes (e.g., boomerang shape as observed in source COALAS-SW.09), "10" for sources whose collapsing contour sizes are obviously large, "01" for collapsing contours comparable to the beam size (i.e., marginally resolved), and "00" to those sources that remain unresolved. In the \underline{Collapsed Size II} classification, we assign ``1'' to sources whose outermost contour of the collapsed image is larger than the synthesized beam size, and "0" to the rest of the unresolved sources.}. Subsequently, in Option V, we simplified the PVD criterion by reducing the bit numbers from two bits (PVD I) to one bit (PVD I) \footnote{In the \underline{PVD I} classification, we assign ``11'' to sources whose PVD exhibits a clear ``S'' shape, indicating a regular velocity gradient, ``10'' to sources with a slight velocity gradient, "01" to sources showing multiple components occupying a wide channel/velocity range (evident from the spectrum aspect with multiple-peak broad spectrum), and "00" for sources with compact components (i.e., no gradient, narrow velocity/channel range) or exhibiting messy PVD behaviours. In the \underline{PVD II}, we combine the first three cases, previously classified as ``11'', ``10'', and ``01'' into a single ``1'', while the rest remain as "0.}. This adjustment aimed to reduce the relative importance of PVD and allocate more weight to the remaining criteria, thus enhancing the overall effectiveness and accuracy of the selection process. In the final Option VII, we introduce an additional constraint to the PVD II classification. Specifically, sources classified as ``1'' in Option VI, but with spectral FWHM below 300 km/s, are re-classified into ``0'' (PVD III) in the Option VII scheme. This constraint is grounded in the rationale that projected extended sources should also exhibit extension along the line-of-sight dimension.

After evaluating various sets of criteria options (Table~\ref{tab:Criteria}), we find that Option VII closely aligns with our requirements, meeting our selection criteria most effectively. As a result, we present Option VII as our final choice for the classification scheme. The detailed description of each criterion is as following:
\begin{table*}
    \centering
    \begin{tabular}{|c|cccccc|}
        \hline
        Option & & & Criteria & & & \\
        \hline
        & 1st & 2nd & 3rd & 4th & 5th & 6th \\
        \hline
        Option I & Collapsed Size I & PVD I & Number SBSP & Fraction SBSP & Fraction E & - \\
        (10 bits) & (2 bits) & (2 bits) & (2 bits) & (2 bits) & (2 bits) & \\
        \hline
        Option II & Collapsed Size I  & PVD I & SNR & Fraction SBSP & Fraction E & - \\
        (9 bits) & (2 bits) & (2 bit) & (1 bits) & (2 bits) & (2 bits) & \\
        \hline
        Option III & PVD I & Collapsed Size I  & Number SBSP & Fraction SBSP & Fraction E & - \\
        (10 bits) & (2 bits) & (2 bits) & (2 bits)& (2 bits) & (2 bits) & \\
        \hline
        Option IV & Collapsed Size I  & PVD I & SNR & Number SBSP & Fraction SBSP & Fraction E \\
        (11 bits) & (2 bits) & (2 bits) & (1 bit) & (2 bits) & (2 bits) & (2 bits)\\
        \hline
        Option V & Collapsed Size II & PVD I & SNR & Number SBSP& Fraction SBSP & Fraction E\\
        (10 bits) &  (1 bit) & (2 bits) &  (1 bit) & (2 bits) & (2 bits) & (2 bits)\\
        \hline
        Option VI & Collapsed Size III & PVD II & SNR & Number SBSP & Fraction SBSP & Fraction E\\
        (9 bits) &  (1 bit) & (1 bits) &  (1 bit) & (2 bits) & (2 bits) & (2 bits)\\
        \hline
        Option VII & Collapsed Size III & PVD III & SNR & Number SBSP& Fraction SBSP & Fraction E\\
        (9 bits) &  (1 bit) & (1 bits) &  (1 bit) & (2 bits) & (2 bits) & (2 bits)\\
        \hline
    \end{tabular}
    \caption{The evolution of criteria options and the final classification scheme for ranking and selecting extended molecular gas reservoirs.}
    \label{tab:Criteria}
    \end{table*}
    
\textbf{Collapsed image size (Collapsed Size I):} assess whether sources are resolved or not via comparing the size of collapsed images with the size of the synthesised beam of the corresponding pointing. The area encompassed by the standard deviation value contour is calculated. Finally, ratios of the collapsed component area and synthesised beam area are calculated. Sources whose ratio is greater than one (i.e., the source size is larger than the synthesised beam) would be classified as resolved. We visually re-checked the collapsed images afterwards, and excluded sources which have multiple contour-encompassed regions separately distributed.

\textbf{Features observed in the position-velocity diagram (PVD I):} evaluate whether sources have multiple components or are velocity gradient on PVDs. An extended source is more likely to be three-dimensional extended rather than only in the projected view. Thus, we applied an extra FWHM-cut at 280 km/s. The binary values of PVDs are determined based on these two factors in a logical conjunction way.

\textbf{{Signal-to-noise ratio (SNR):}} SNR values are from \citet{Jin_2021AA...652A..11J}. We assign the bit value to sources whose SNR values are greater than 5.0 and assign zeros to the others.

The following three criteria are about the observational conditions, and they are relevant to observational pointings of each source. The ATCA survey of the Spiderweb protocluster in total has 13 pointings, and each of the 46 CO emitters are covered by one up to five ATCA observational pointings.

\textbf{The number of ATCA observational pointings covering the source with a small beam size ($<$7$\farcs$0) (Number of Small-Beam-Size Pointings - SBSP):} the sources are more likely to be resolved with observational pointings of smaller beam size. Two bits are employed and classify the cases into four classes: ``11'' for sources covered by three or four small beam size pointings, ``10'' for two, ``01'' for one, and ``00'' for none. 

\textbf{The fraction of small-beam-size pointings ($<$7$\farcs$0) relative to the total number of pointings that cover the source (Fraction of Small Beam Size Pointings - Fraction SBSP):} two bits are employed and the cases are divided into four classes:
\begin{enumerate}
    \item ``11'': fraction value equal to 1.0 (1/1, 2/2, 3/3, 4/4, 5/5; all the pointings covering this source have small beams);
    \item ``10'': fraction values equal to 0.6 (3/5), 0.67 (2/3), 0.75 (3/4), and 0.8 (4/5) (more than half pointings have small beams); 
    \item ``01'': fraction values equal to 0.2 (1/5), 0.25 (1/4), 0.33 (1/3), 0.4 (2/5), and 0.5 (1/2, 2/4) (fewer than half pointings have small beams);
    \item ``00'': fraction values equal to 0.0 (0/1, 0/2, 0/3, 0/4, 0/5; none of the pointings cover this source have a small beam)
\end{enumerate}

\textbf{The fraction of pointings within which the source is located at the edge (Fraction E):} sources located at the edge of a single-pointing may be subject to significantly higher noise levels, which can markedly impact the accuracy of the ranking results. We calculate the distances between sources and the pointing centre and classify the source as located at the edge of observational pointings if the distance is greater than 94$\%$ of the primary beam size radius. We use two bits for this criterion:
\begin{enumerate}
    \item ``11'': fraction values equal to 0.0 (0/1, 0/2, 0/3, 0/4, 0/5; within none of the pointings covering this source, it is located at the edge of the pointing)
    \item ``10'': fraction values equal to 0.2 (1/5), 0.25 (1/4), 0.33 (1/3), 0.4 (2/5), and 0.5 (1/2, 2/4) (within fewer than half of the pointings covering this source, it is located at the edge of the pointing);
    \item ``01'': fraction values equal to 0.6 (3/5), 0.67 (2/3), 0.75 (3/4), and 0.8 (4/5) within more than half of the pointings covering this source, it is located at the edge of the pointing; 
    \item ``00'': fraction values equal to 1.0 (1/1, 2/2, 3/3, 4/4, 5/5; within all of the pointings covering this source, it is located at the edge of the pointing);
\end{enumerate}

\subsection{Design of the binary criteria ranking: the boundary ``score'' between candidates and the others}
For better understanding, please refer to Figure A2 while reading the following explanations of the decimal separation line of "400": 
\label{sec:DesignSeparationLine}
\begin{enumerate}
\item If a source is assigned with highest marks for each criterion (the maximum case in equation~\ref{equ:maxmin}: all the $x_i$=1, and the corresponding decimal value maximised to 511), this source would certainly be classified as an extended gas candidate;
    \item The case which fails either the ``Size of collapsed images'' (the first criteria; fail: bit value ``0'') or ``Position Velocity Diagram'' (the second criteria; fail: bit value ``0'') would not reach the boundary score ``400'' as the following equation shows:
    \begin{equation}
        \text { Score }=\sum_{i=0}^{N-1} x_i \cdot 2^i \stackrel{N=9}{===}\left\{\begin{array}{l}
        \leq\;255\;(if \; x_8=0; Collapsed\;Size) \\
        \leq\;383\;(if \; x_7=0; PVD)
        \end{array}\right.
    \end{equation}
    \item Assuming that a source meets the condition of fulfilling the first two criteria (i.e., both criteria ``Size of collapsed images'' and ``Position Velocity Diagram'' are assigned bit number ``1''), the third criteria, SNR, could be either ``1'' or ``0'' to be above the boundary score:
    \begin{enumerate}
        \item SNR = ``1'', together with two ``1'' from the first two criteria, will result in a ``score'' greater or equal to 448: 
        \begin{equation}
            \text { Score }=\sum_{i=0}^{N-1} x_i \cdot 2^i \stackrel{N=9}{\geq}448\;(if \; x_8, x_7, x_6=1) 
        \end{equation}
        \item SNR = ``0'' requires the binary numbers of the fourth criteria, Number SBSP, to be greater than ``00'', i.e., the source is at least covered by one pointing with a small beam (i.e., relatively high resolution).
        \begin{enumerate}
            \item Number SBSP = ``11'' would result in a decimal value greater or equal to 432;
            \item Number SBSP = ``10'' would result in a decimal value greater or equal to 416;
            \item Number SBSP = ``01'' would result in a decimal value greater or equal to 400.
        \end{enumerate}
    \end{enumerate}
\end{enumerate}

\begin{figure*}
    \centering
    \includegraphics[angle=90, width = 0.65\linewidth]{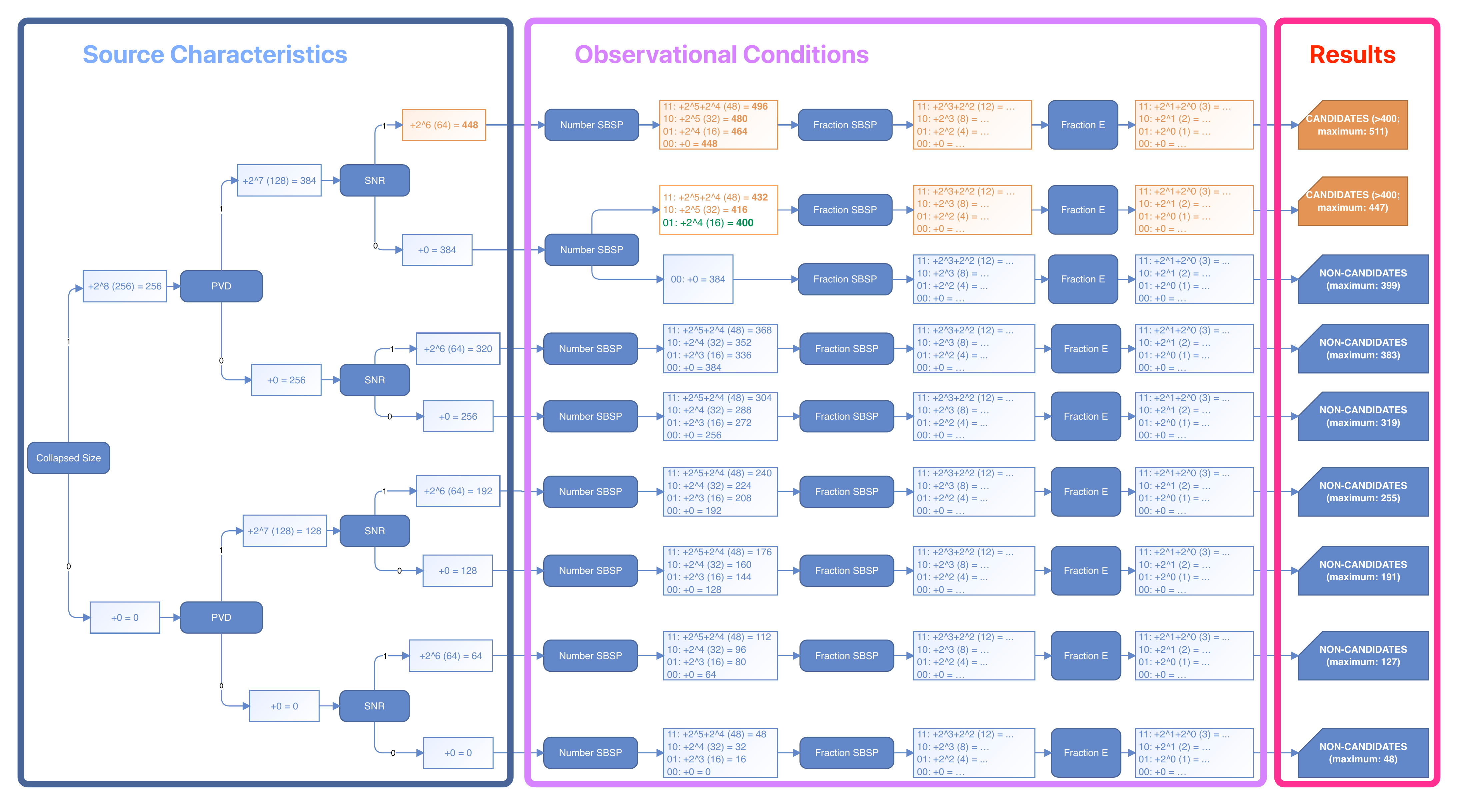}
    \caption{Visualisation of the Criteria Ranking Bit Assignments, Calculation of the Ranking "Score", and Selection of Large Gas Reservoir Candidates. The dark blue boxes represent abbreviations of the six criteria, with higher priority shown on the left and lower priority on the right. The first three criteria (Collapsed sizes, PVDs, and SNRs), enclosed within the large blue box, depict the source characteristics, while the later three criteria (Number SBSP, Fraction SBSP, and Fraction E), enclosed within the purple box, are related to the observational conditions. The red box highlights the results of candidate selection, with selected candidates in orange and non-selected candidates in blue.}
    \label{fig:CriteriaRanking__}
\end{figure*}

In the cases explained above, the first four criteria would be enough to determine whether a source could be a large gas reservoir candidate or not. The lower limit for the selection of candidates is the combination [Collapsed size ``1''] + [PVD ``1''] + [SNR ``0''] + [Number SBSP `01''] + [Fraction SBSP ``--''] + [Fraction E ``--''] (Figure~\ref{fig:CriteriaRanking__}). 

\section{Criteria Calibrators}
\label{sec:A_CriteriaCalibrators}
The information of seven calibrators are summarised in Table~\ref{tab:calibrators}, including the source ID, alias, reference paper, data available, and corresponding figures of collapsed images and channel maps.

\begin{table*}
    \centering
    \begin{tabular}{c|c|c|c|c}
        \hline\hline
        Calibrators & Alias &  Reference(extended) / Notes & Data &  Figure  \\
        \hline\hline
        COALAS-SW.01 & HAE229 & \citet{Dannerbauer_2017AA...608A..48D} & High-resolution; mosaic & - \\
        \hline
        COALAS-SW.02 &  Spiderweb galaxy & \citet{Emonts_2016Sci...354.1128E, Emonts_2018MNRAS.477L..60E} & High-resolution; mosaic & - \\
        \hline
        COALAS-SW.03 & DKB03 & This work: extended & High-resolution; mosaic  & {Figure~\ref{fig:DKB03_Mosaic_AND_HighResolution},\ref{fig:DKB03_ChannelMap}}\\
        \hline
        COALAS-SW.06 & - & This work: extended & High-resolution; mosaic  &  Figure~\ref{fig:COALAS_06} \\
        \hline\hline
        COALAS-SW.29 & - & This work: merger/rotating-like & mosaic & Figure~\ref{fig:COALAS_29_ATCA} \\
        \hline
        COALAS-SW.23 & - & This work: sharing giant molecular gas with COALAS-SW.46 & mosaic & Figure~\ref{fig:COALAS_23_46_ATCA} \\
        \hline
        COALAS-SW.46 & - & This work: sharing giant molecular gas with COALAS-SW.23 & mosaic & Figure~\ref{fig:COALAS_23_46_ATCA} \\
        \hline\hline
    \end{tabular}
    \caption{Seven calibrators of the presented method. The first four are  confirmed large gas reservoirs confirmed through high-resolution observations, and the final three have strong indicators to be large gas reservoirs with solely mosaic data.}
    \label{tab:calibrators}
\end{table*}
\begin{figure*}
    \centering
    \includegraphics[width = 1.0\linewidth]{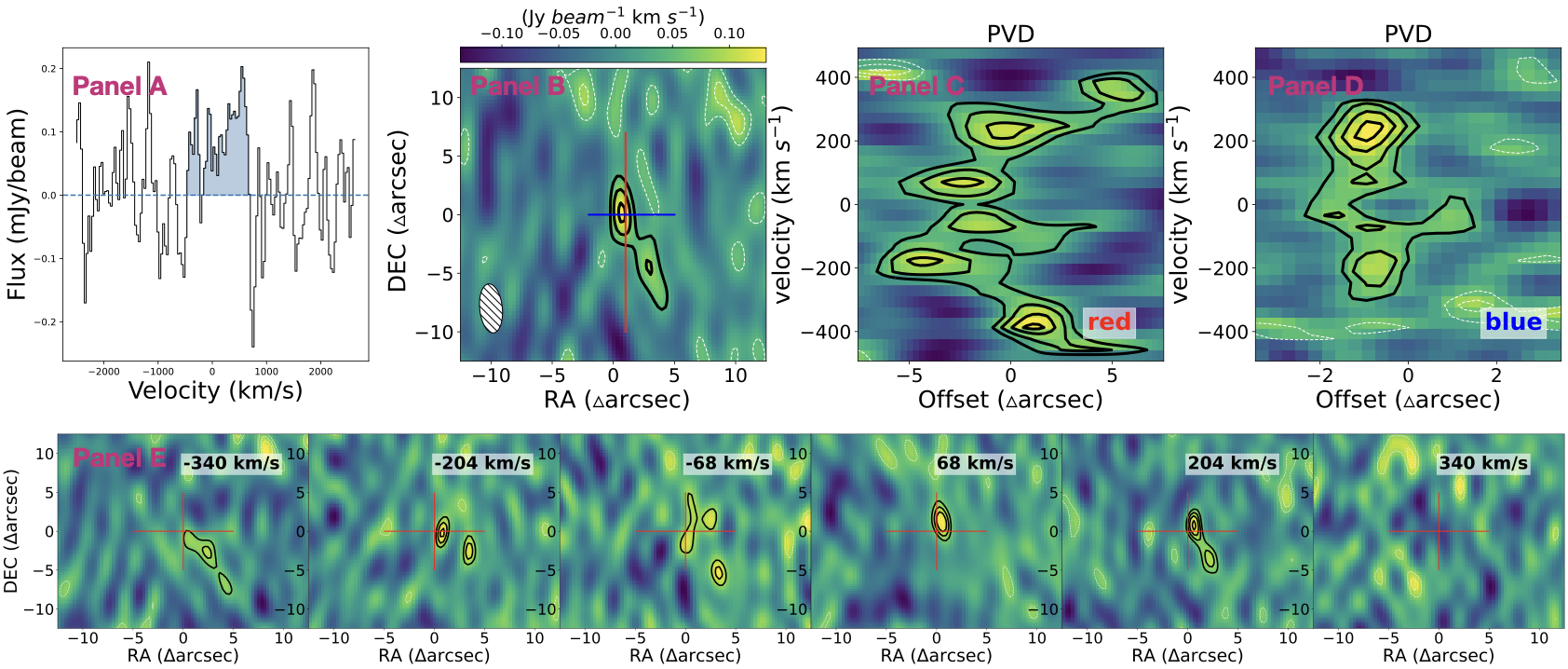}
    \caption{High-resolution plots of COALAS-SW.06. Panel A is the spectrum extracted from the emission peak with aperture size of 1$\farcs$5. Panel B is the moment 0 map of COALAS-SW.06, Panel C and Panel D are PVDs extracted along red and cyan lines shown in Panel B, and Panel E presented channel maps. The contour levels in the collapsed images are presented at [1, 2, 3] times the standard deviation values, with the source delineated by black solid lines, and noise indicated by white dashed lines.}
    \label{fig:COALAS_06}
\end{figure*}
\begin{figure*}
    \centering
    \includegraphics[width = 1.0\linewidth]{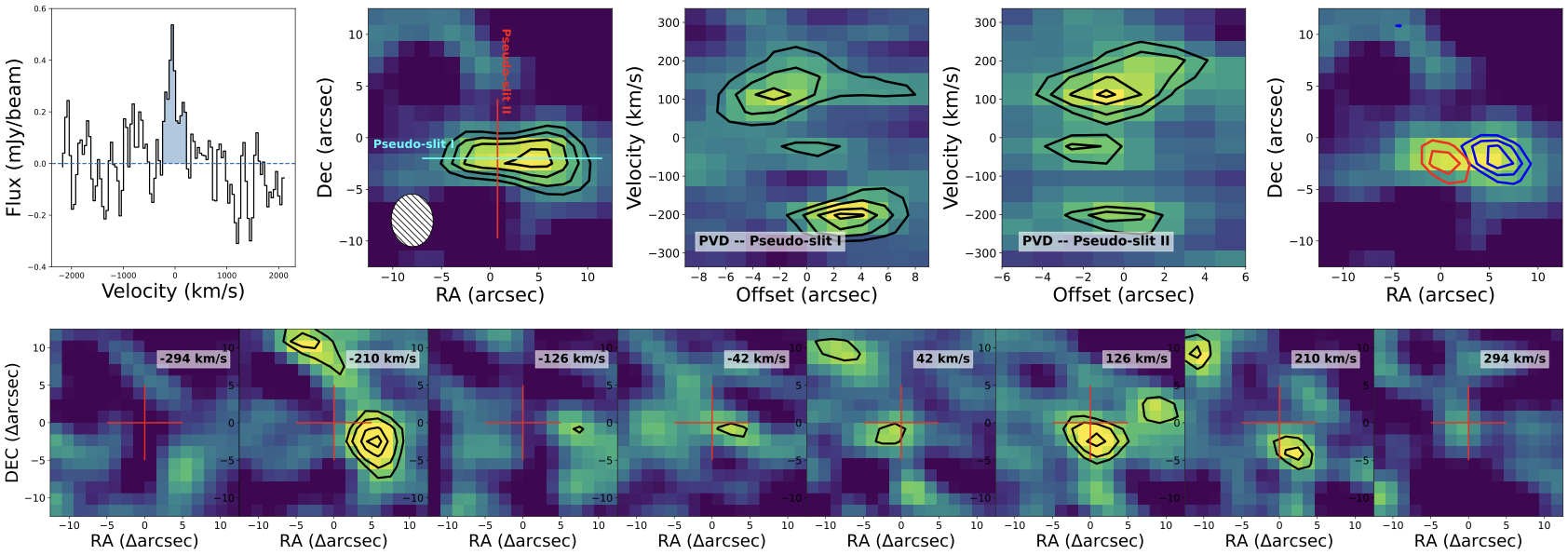}
    \caption{Characterise COALAS-SW.29 based on mosaic data. The upper left panel is the spectrum of COALAS-SW.29. The following four panels are collapsed image with black contours representing significance levels of [2.0, 2.5, 3.0, 3.5] times the standard deviation ($\sigma$ = 0.033 Jy $\rm {beam}^{-1}$ km $\rm s^{-1}$), PVDs extracted along the cyan and red lines, and contours of collapsed images from bluer and redder channels (corresponding to the negative and positive velocity components on PVDs). The second row shows channel maps of extended CO(1-0).}
    \label{fig:COALAS_29_ATCA}
\end{figure*}
\begin{figure*}
    \centering
    \includegraphics[width = 1.0\linewidth]{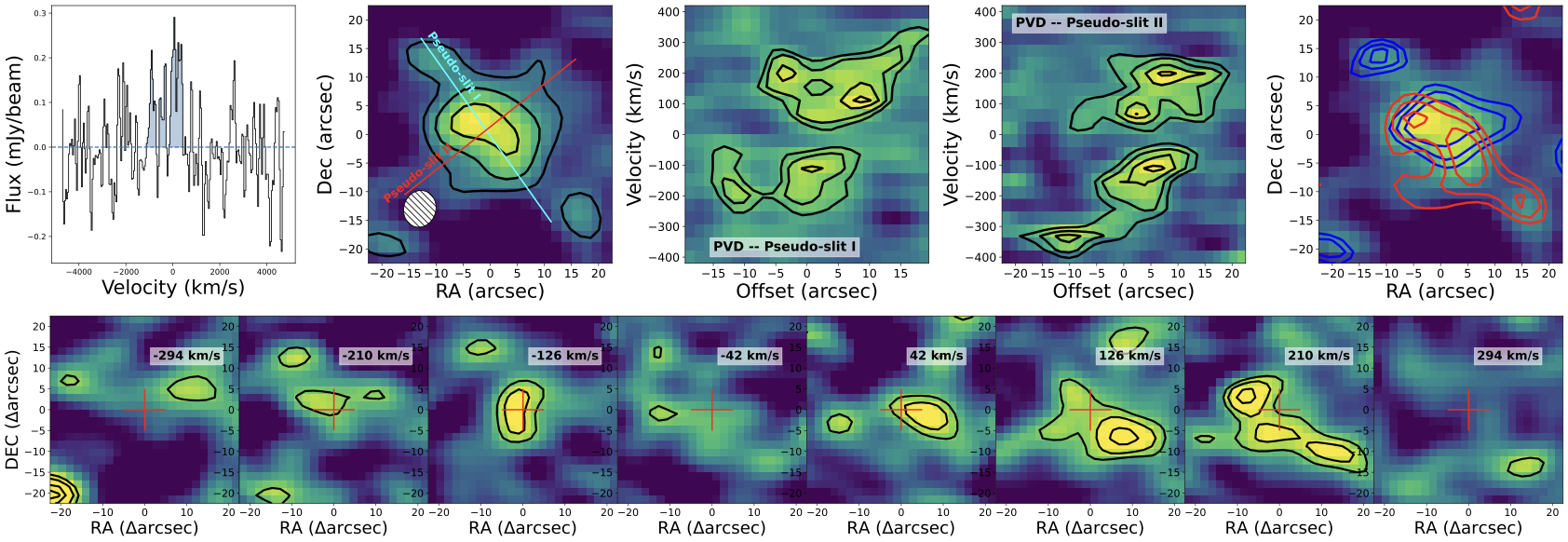}
    \caption{Characterise COALAS-SW.23 and COALAS-SW.46 based on mosaic data. The upper left panel is the spectrum of COALAS-SW.23. The following four panels are collapsed image with black contours representing significance levels of [1.5, 2.5, 3.5, 4.5] times the standard deviation ($\sigma$ = 0.028 Jy $\rm {beam}^{-1}$ km $\rm s^{-1}$), PVDs extracted along the cyan and red lines, and contours of collapsed images from bluer and redder channels (corresponding to the negative and positive velocity components on PVDs). The second row shows channel maps of extended CO(1-0).}
    \label{fig:COALAS_23_46_ATCA}
\end{figure*}
\section{Robust Candidates}
\label{sec:A1_Robust}
The collapsed images and PVDs of the 14 robust candidates are presented in Figure~\ref{fig:FinalCandidatesPlot_I}.

\begin{figure*}
    \centering
    \includegraphics[width = 0.75\linewidth]{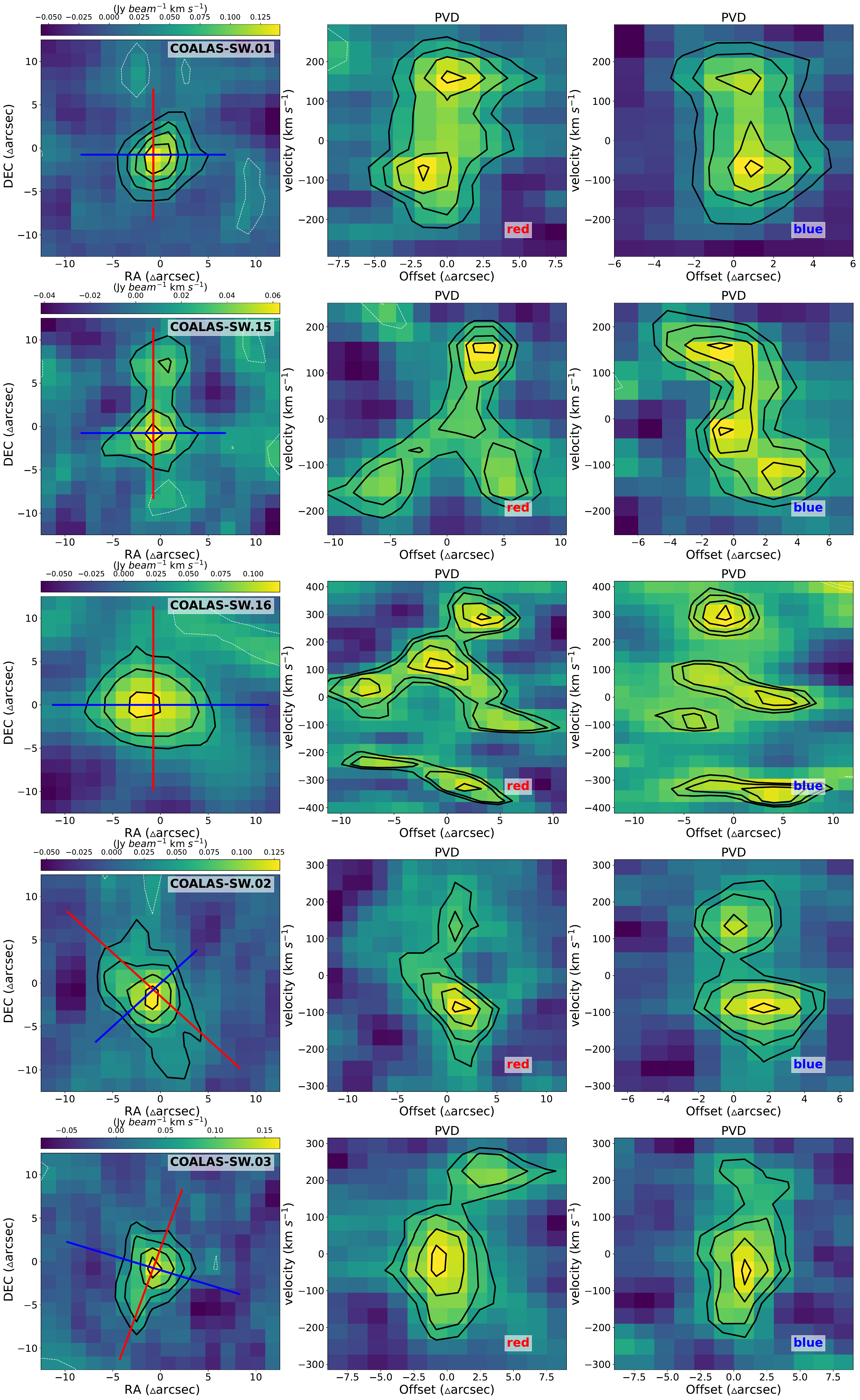}
    \caption{The figure displays robust candidates in individual rows. The leftmost panels show the collapsed images, while the middle and right panels present PVDs extracted from two perpendicular pseudo-slits. The source ID for each robust candidate is indicated in the upper right corner of the corresponding collapsed image. The contour levels in the collapsed images are presented at [1, 2, 3, 4] times the standard deviation values. The emission contours are depicted in black solid lines, while the noise are represented by white dashed lines.}
    \label{fig:FinalCandidatesPlot_I}
\end{figure*}
\begin{figure*}
    \setcounter{figure}{0}
    \centering
    \includegraphics[width = 0.78\linewidth]{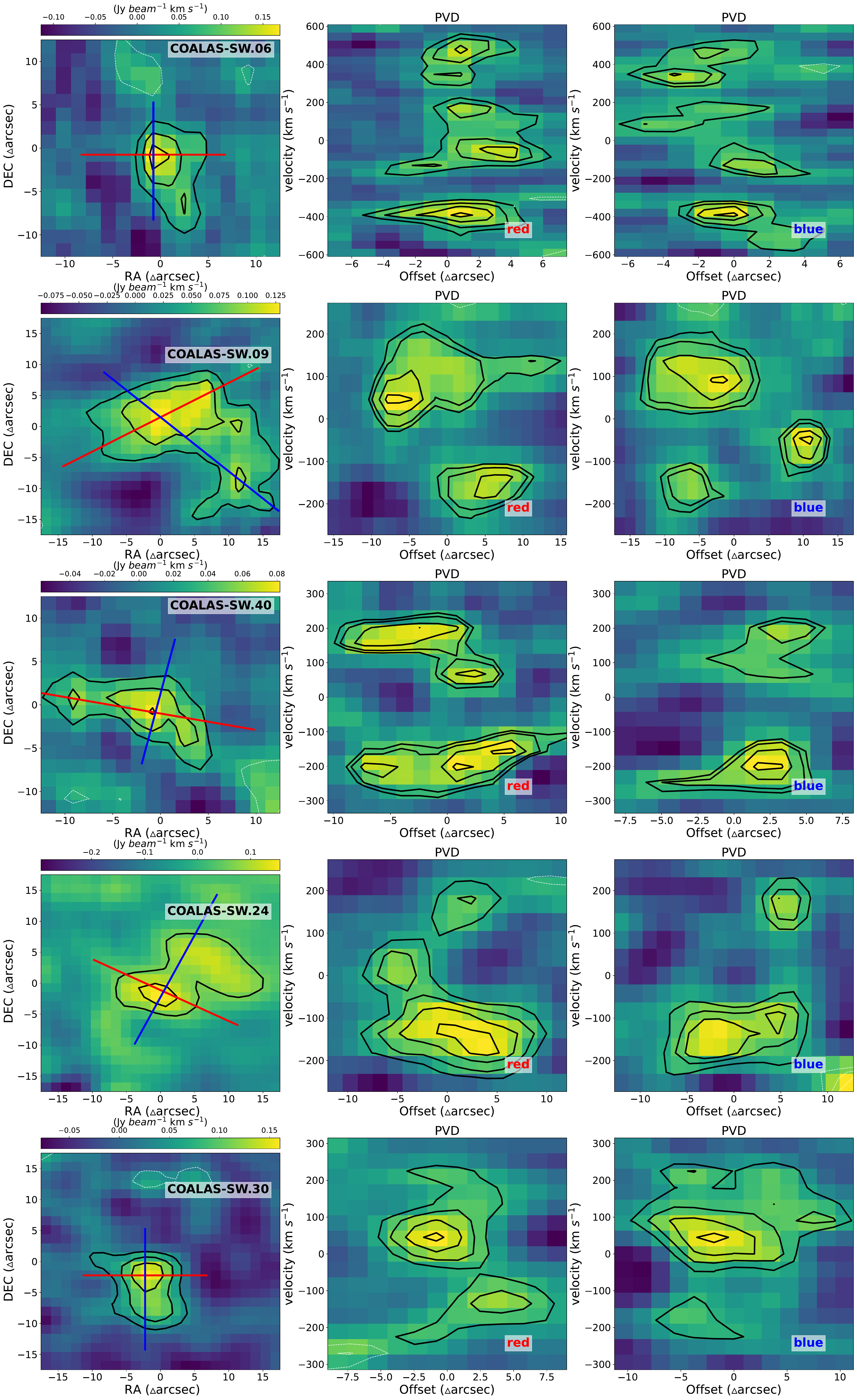}
    \caption{(Continued.)}
    \label{fig:FinalCandidatesPlot_II}
\end{figure*}
\begin{figure*}
    \setcounter{figure}{0}
    \centering
    \includegraphics[width = 0.85\linewidth]{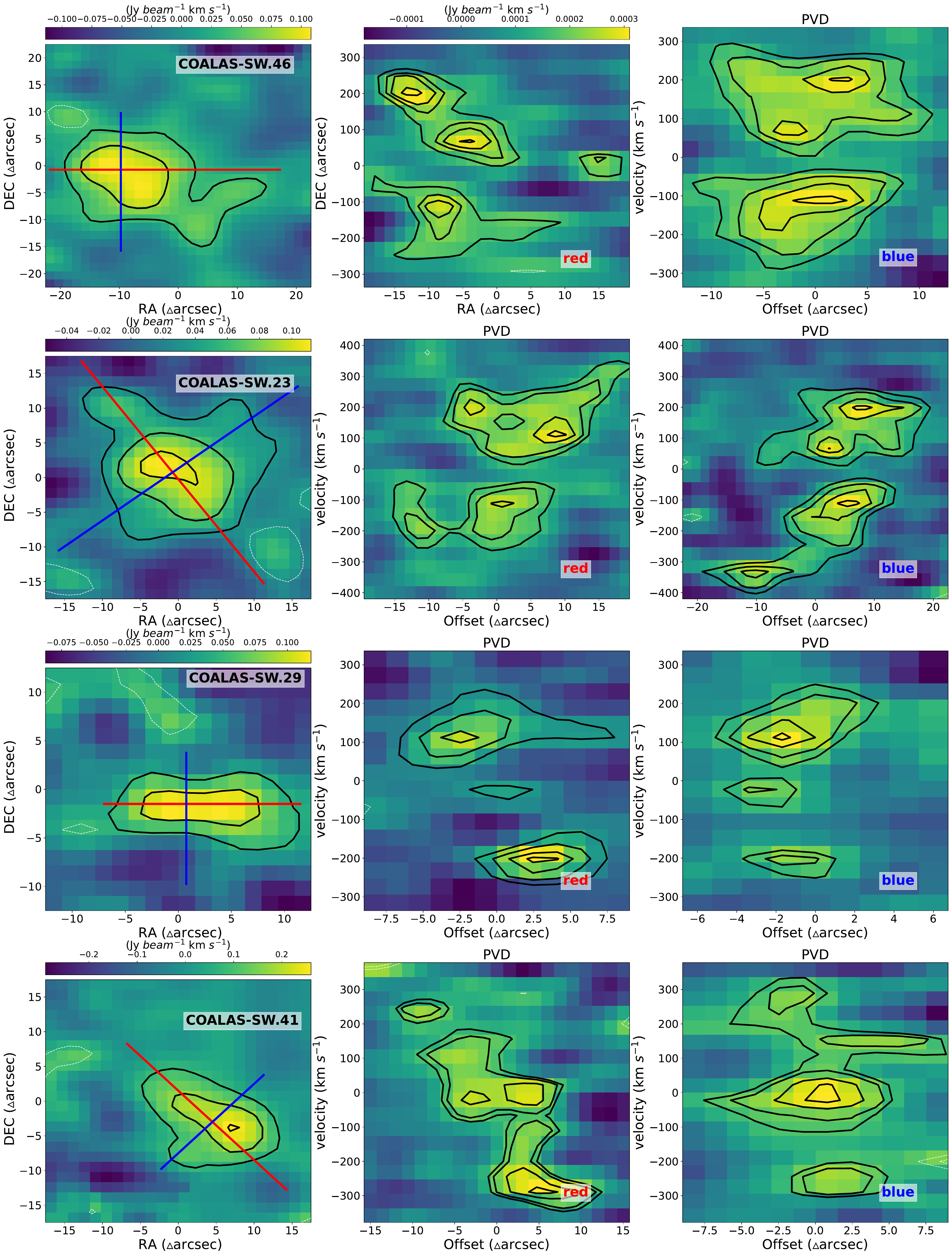}
    \caption{(Continued.)}
    \label{fig:FinalCandidatesPlot_III}
\end{figure*}
\section{Tentative Candidates}
\label{sec:A2_Tentative}
The collapsed images and PVDs of the seven tentative candidates are presented in Figure~\ref{fig:FinalCandidatesPlot_IV}.
\begin{figure*}
    \centering
    \includegraphics[width = 0.75\linewidth]{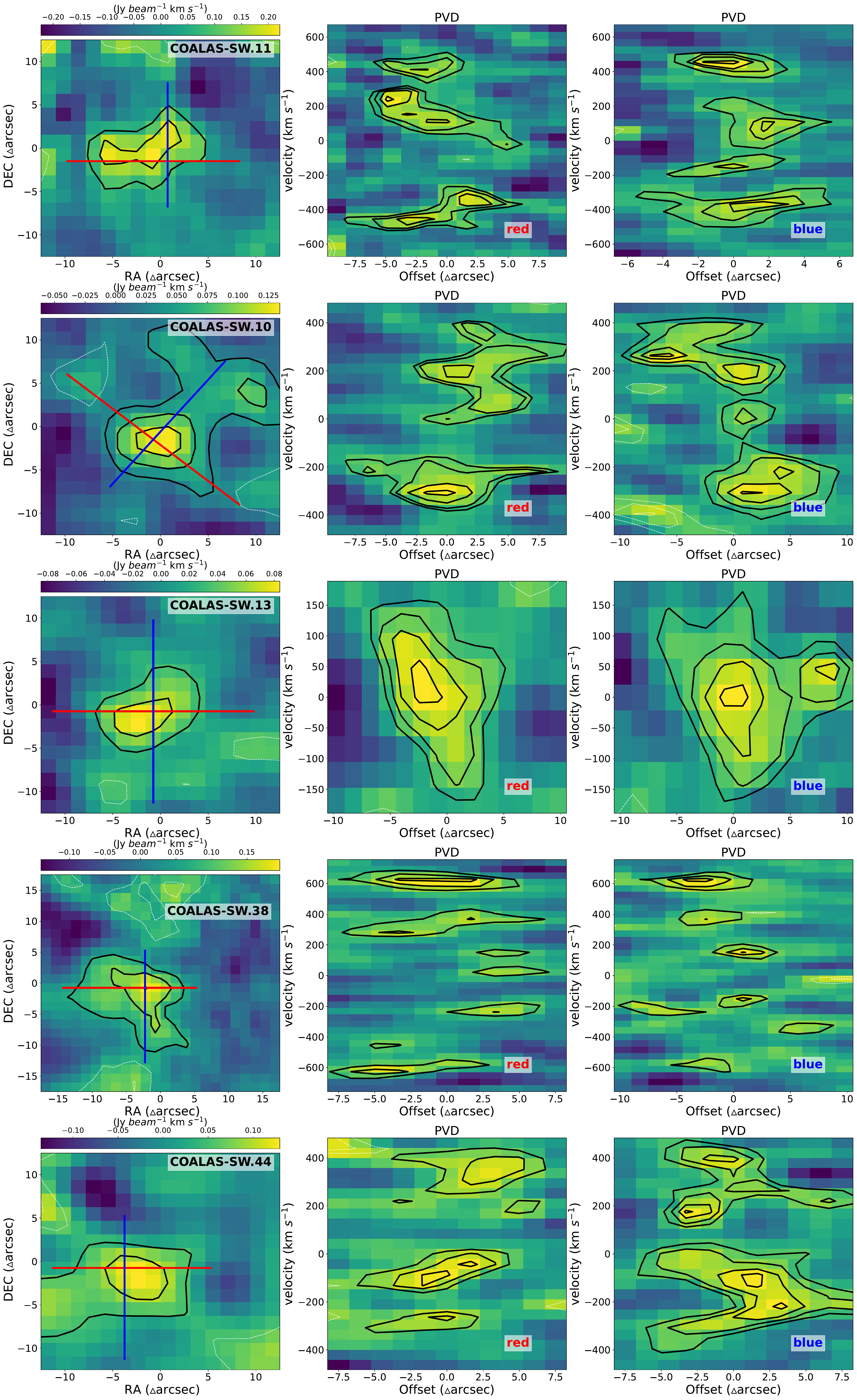}
    \caption{Tentative candidates. Same scheme as Figure~\ref{fig:FinalCandidatesPlot_I}}
    \label{fig:FinalCandidatesPlot_IV}
\end{figure*}
\begin{figure*}
    \setcounter{figure}{0}
    \centering
    \includegraphics[width = 0.8\linewidth]{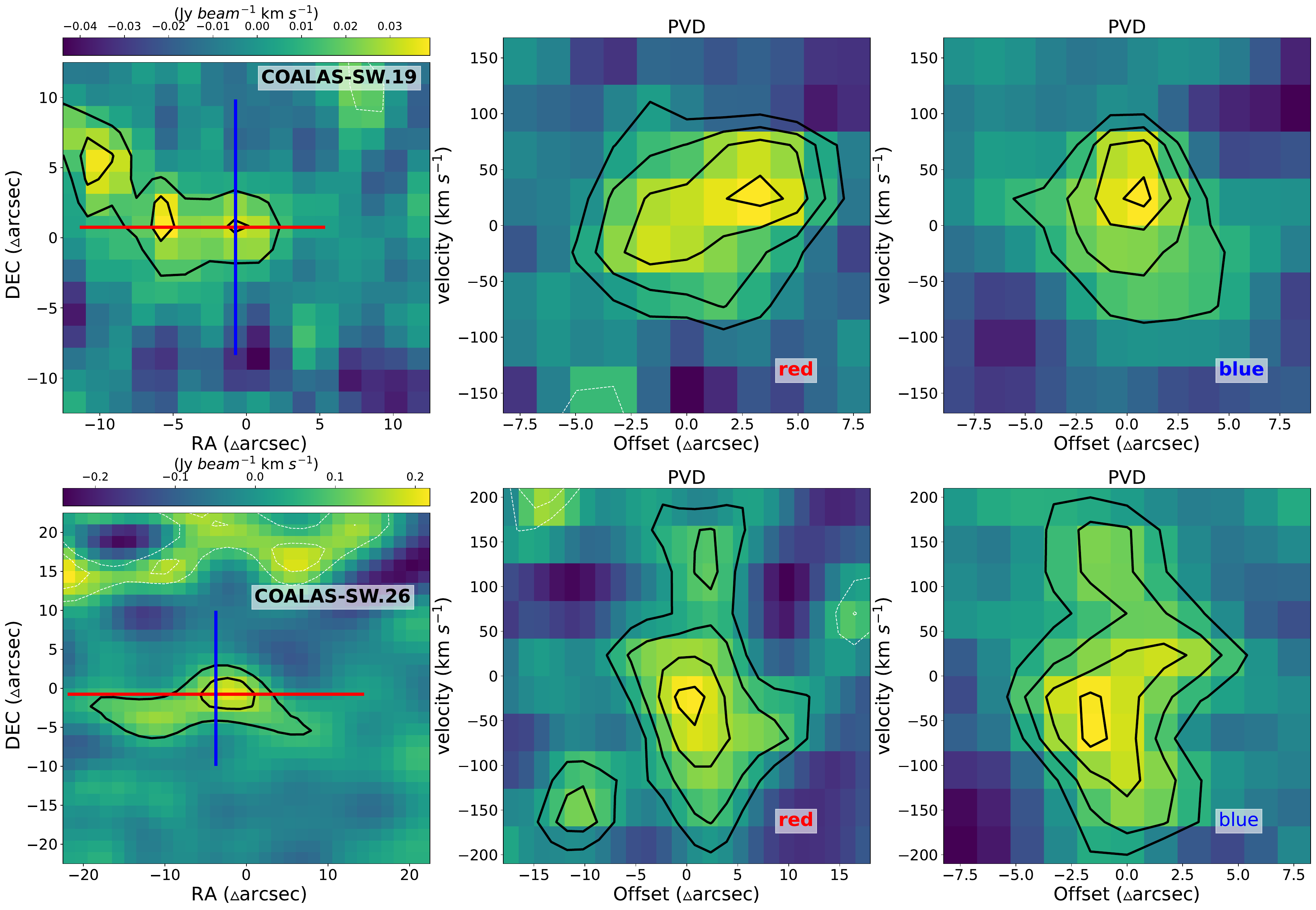}
    \caption{(Continued.)}
    \label{fig:FinalCandidatesPlot_V}
\end{figure*}

\section{Statistical explanation}
For the contours of collapsed images in Figure~\ref{fig:FinalCandidatesPlot_I} $\&$ Figure~\ref{fig:FinalCandidatesPlot_IV}, we employed the STandard Deviation (STD) values rather than the commonly used Root Mean Square (RMS) values. 

The equations of RMS is 
\begin{equation}
RMS=\sqrt{\frac{1}{n} \sum_i x_i^2}
\end{equation}
where n is number of values, and $x_i$ is each values. 

The equation of standard deviation is
\begin{equation}
STD=\sqrt{\frac{1}{n}\sum\left(x_i-\mu\right)^2}
\end{equation}
where n is the number of values, $x_i$ is each values, and $\mu$ is the mean value. In physics, the term root mean square is used as a synonym for standard deviation when it can be assumed that the input number population has a zero mean, i.e., referring to the square root of the mean deviation of a signal from a given baseline or fit.

\begin{figure*}
    \centering
    \includegraphics[width = 0.8\linewidth]{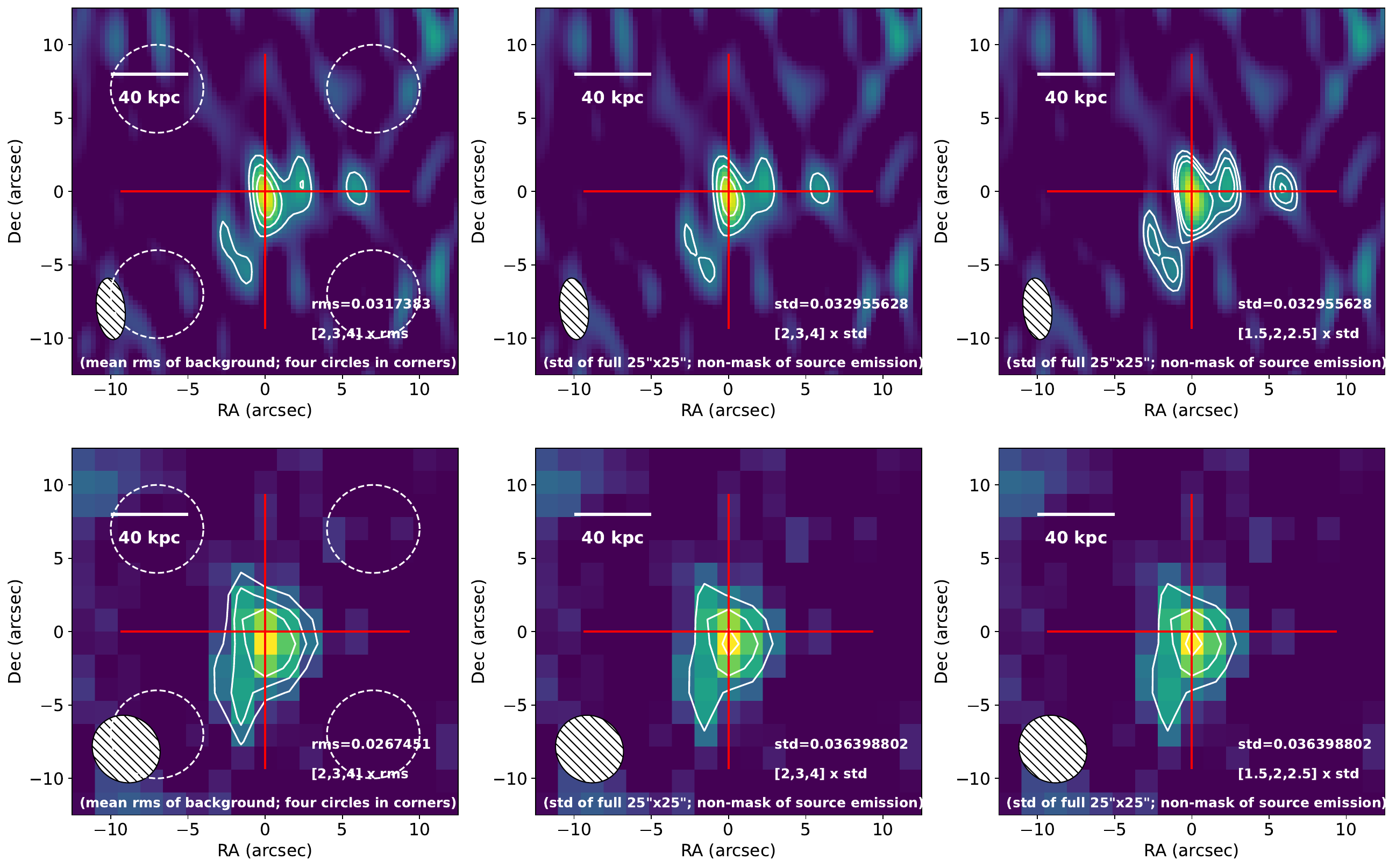}
    \caption{Comparison between the usage of Root Mean Square (RMS) values by STandard Deviation (STD) values. Upper panels are collapsed images based on the high-resolution data of COALAS-SW.03, and bottom panels are based on the mosaic/coarse data. In the left columns the overlaid contours are [2, 3, 4] $\times$ RMS (RMS values are calculated based on four regions with radius of 3$\farcs$0 marked with dashed white circles). In the middle columns, the contours are [2, 3, 4] $\times$ STD, in the right columns [1.5, 2.0, 2.5] $\times$ STD.}
    \label{fig:RMS_STD}
\end{figure*}

The reason and feasibility are explained here, taking COALAS-SW.03 as an example. First, we manually selected four regions with the radius of 3$\farcs$0, obtained the RMS values, and overlaid the [2, 3, 4] $\times$ RMS contours on the collapsed images as shown in the left column panels in Figure~\ref{fig:RMS_STD}. If we apply this method to all the other sources, we need to be careful when we manually select the regions for such background RMS calculations. If the regions are too large, they might be contaminated by the source emission and result in a larger RMS value; if the regions are too small, this might not be representative; if the positions of regions are inappropriate, the results might be biassed by some regions of extreme values. The spatial scales of emission vary among our sample, and thus requires plenty of manual work on RMS calculation as explained above. Second, to simplify the work, we tried the contour levels based on standard deviation values. We calculated the standard deviation values based on these 25$\farcs$0 $\times$ 25$\farcs$0 cutout without masking the emission from the central source. Keeping the level times the numbers of [2, 3, 4], we obtained the contour plots in the middle column in Figure~\ref{fig:RMS_STD}, whose outermost contour is more conservative/smaller compared to the panels of the left column. It is reasonable as we included the source emission when calculating the STD values and these derived STD values are greater than the RMS values obtained from the background regions. To get similar contours, we tried the smaller level times [1.5, 2.0, 2.5], and displayed them in the right panels of Figure~\ref{fig:RMS_STD}. To conclude, with a lower multiple times, we can use the standard deviation values derived without masking the source to replace the calculations of RMS values requiring lots of manual work. To conclude, for simplifying the work on presenting the large gas reservoirs, we employed the [1, 2, 3, 4] $\times$ STD values for collapsed images in Figure~\ref{fig:FinalCandidatesPlot_I} and Figure~\ref{fig:FinalCandidatesPlot_IV}.


\label{lastpage}
\end{document}